\documentclass[aip,preprint]{revtex4-2}
\usepackage[cp1251]{inputenc}
\usepackage[T2A]{fontenc}
\usepackage[english]{babel}
\usepackage{amssymb,latexsym,amsmath,amscd}
\usepackage{graphicx,color,framed}
\usepackage{bm}
\usepackage{tikz}
\usepackage{xcolor}
\usepackage{siunitx}
\usepackage{empheq}
\usepackage{subcaption}
\usepackage{xr}
\usepackage{microtype}
\usepackage{amsfonts}
\usepackage{pgfplots}
\pgfplotsset{width=10cm,compat=1.9}
\usepackage{pgfplotstable}
\usepackage[normalem]{ulem}
\graphicspath{{figures/}}

\makeatletter
\newcommand*{\addFileDependency}[1]{
  \typeout{(#1)}
  \@addtofilelist{#1}
  \IfFileExists{#1}{}{\typeout{No file #1.}}
}
\makeatother

\begin{document}
\pgfplotsset{every axis/.append style={line width=1.3pt}}
\tikzset{every mark/.append style={scale=1.2}}
\affiliation{Laboratory of Computational Physics, HSE University, Tallinskaya st. 34, 123458 Moscow, Russia}
\author{\firstname{Yury A.} \surname{Budkov}}
\affiliation{Laboratory of Computational Physics, HSE University, Tallinskaya st. 34, 123458 Moscow, Russia}
\affiliation{Laboratory of Multiscale Modeling of Molecular Systems, G.A. Krestov Institute of Solution Chemistry of the Russian Academy of Sciences, 153045, Akademicheskaya st. 1, Ivanovo, Russia}
\affiliation{A.N. Frumkin Institute of Physical Chemistry and Electrochemistry, Russian Academy of Science 119071, 31 Leninsky Prospect, Moscow, Russia}
\email[]{ybudkov@hse.ru}
\author{\firstname{Nikolai N.} \surname{Kalikin}}
\affiliation{Laboratory of Computational Physics, HSE University, Tallinskaya st. 34, 123458 Moscow, Russia}
\affiliation{Laboratory of Multiscale Modeling of Molecular Systems, G.A. Krestov Institute of Solution Chemistry of the Russian Academy of Sciences, 153045, Akademicheskaya st. 1, Ivanovo, Russia}
\author{\firstname{Petr E.} \surname{Brandyshev}}
\affiliation{Laboratory of Computational Physics, HSE University, Tallinskaya st. 34, 123458 Moscow, Russia}

\title{Surface Tension of Aqueous Electrolyte Solutions. A Thermomechanical Approach}

\begin{abstract}
We determine the surface tension of aqueous electrolyte solutions in contact with non-polar dielectric media using a {\sl thermomechanical} approach, which involves deriving the stress tensor from the thermodynamic potential of an inhomogeneous fluid. To obtain the surface tension, we calculate both the normal and tangential pressures using the components of the stress tensor, recently derived by us [Y. A. Budkov and P. E. Brandyshev, The Journal of Chemical Physics 159 (2023)] within the framework of Wang's variational field theory. Using this approach, we derive an analytical expression for the surface tension in the linear approximation. At low ionic concentrations, this expression represents the classical Onsager-Samaras limiting law. By utilizing only one fitting parameter, which is related to the affinity of anions to the dielectric boundary, we can approximate various experimental data regarding the surface tension of aqueous electrolyte solutions. This approximation applies to both the solution-air and solution-dodecane interfaces, covering a wide range of electrolyte concentrations.

\end{abstract}

\maketitle

\section{Introduction}
Since the early twentieth century, it has been recognized that the dissolution of most inorganic salts in water leads to an increase in the surface tension of the solution. Wagner \cite{wagner1924oberflachenspannung} first formulated the theory of the surface tension of strong electrolyte solutions for the case of electrolyte solution-gas interface, which was later expanded in the work of Onsager and Samaras, based on the Debye-H{\"u}ckel theory foundations. However, the Onsager-Samaras (OS) theory focused on the repulsive forces experienced by an ion in a solution near a flat boundary separating two continuous dielectric media~\cite{onsager1934surface}. These forces, known as "image forces"~\cite{landau2013electrodynamics}, depend only on the charge of the ion and are not affected by other characteristics. According to this theory, the surface tensions of aqueous solutions containing a symmetric 1:1 electrolyte at a certain concentration should be the same, and both cations and anions should contribute equally to the adsorption. However, measurements of the electrolyte solutions surface tension across a wide range of concentrations did not align with this theory.

Extensive research employing molecular dynamic simulations, quantum chemistry calculations, and vibrational spectroscopy investigation has provided valuable insights regarding the molecular structure of the aqueous electrolyte-air interface (see, for instance,~\cite{mucha2005unified,fennell2009ion,jungwirth2002ions}). These studies have revealed that variations in such properties as polarizability, charge magnitude, and ion size play a crucial role in shaping the charge distribution and electrical potential profile within the surface layer of the solution. Notably, while monovalent and divalent cations of inorganic salt are repelled from the solution's surface, anions tend to accumulate within the surface layer. Consequently, the degree of adsorption exhibited by these anions is augmented as their polarizability increases. From general considerations, the latter effect should affect the surface tension at rather large electrolyte concentrations exhibiting the effect of ion specificity, disregarded within seminal OS theory.

Shchekin and Borisov~\cite{shchekin2005thermodynamics} were the first who attempted to address the limitations of the OS theory by incorporating the insights from Ninham and colleagues~\cite{ninham1997ion,ninham1997hydrophobicity}. In addition to the account of the electrostatic image forces acting on the ions, the authors also considered the van der Waals forces and structural forces, experienced by the ions near the interface. These forces arise, correspondingly, from the polarizability of the ions and the change in their hydration energy due to the deviation of the solvent density profile at surface from its bulk constant value. This expanded perspective enabled the derivation of an expression for surface tension that appropriately accounts for the specific adsorption of ions at the interface. With this newfound expression, the authors successfully approximated the dependence of the surface tension at the aqueous electrolyte-air interface on electrolyte concentration for several salts up to a concentration of 1 M. The papers by Levin and coworkers~\cite{levin2009polarizable,levin2009ions,dos2010surface,dos2012ions} present the calculation of the solvation free-energy of polarizable ions at air/water and oil/water interfaces. Their approach builds on the standard Poisson-Boltzmann (PB) theory, but with modifications, allowing them to take into account the ion-surface interaction. These modifications involve incorporating various ion-surface interaction terms into the PB equation. These terms model the image charge interaction, Stern exclusion layer, ionic cavitation energy, and ionic polarizability. Levin and colleagues successfully numerically computed the surface tension for a range of sodium salts. They further enhanced their predictions by fitting them to the Hofmeister series. To optimize their calculations, they incorporated the hydration radius of the sodium ion as a single fitting parameter. To apply their theory to different interfaces, one has to incorporate a second fit parameter to adjust the dispersion forces at the oil/water interface.

While several attempts have been made to generalize the OS theory, the self-consistent theoretical framework remained unknown. In other words, the OS theory utilizes the (linearized) PB equation when a planar dielectric boundary is present to derive the one-dimensional image charge potential of mean force. However, it should be noted that this potential is not a solution of the one-dimensional PB equation itself. In fact, the PB solution alone cannot account for any image charge effects. This issue was addressed by Markovich et al.\cite{markovich2014surface,markovich2015surface}, who developed a statistical field theory of electrolyte solutions at a flat dielectric boundary by considering the specific adsorption of ions at the interface. The authors employed a loop expansion method to expand the free energy and included only one loop correction to the mean-field approximation. It was able to reproduce the OS result and the corrections related to the specific adsorption of ions at the interface. Considering only specific adsorption of anions and utilizing one fitting parameter to describe the affinity of anions to the interface, they successfully matched the experimental dependencies of surface tension on concentration for various electrolyte solutions. Essentially, their self-consistent approach yields an analytical prediction that reconciles the OS result with the ionic specificity of the Hofmeister series. Lau and Sokoloff~\cite{Lau2020} have developed a related statistical field theory that not only reproduces the OS result at the one-loop approximation level, but also offers valuable insights into the underlying nature of the image force potential. Specifically, the authors showed that this potential can be derived from the renormalised (or "dressed") by the presence of the ions Green function. The same conclusion was reached by the authors of the papers~\cite{hatlo2008role,wang2016inhomogeneous}, as well.

It is also instructive to discuss two related papers, where surface tension of electrolyte solutions has been systematically analysed in general physical context. Dean and Horgan~\cite{dean2003weak} conducted a comprehensive analysis to determine the surface tension of ionic solutions. They utilized a systematic cumulant expansion method, where the first order calculation is equivalent to the Debye-H{\"u}ckel linear theory. In their model, the interactions between ions and the surface are represented by an ionic exclusion layer with a defined thickness. This approach considers only the length scale of the ion-interface interaction, disregarding any energy scale. The researchers solved this model utilizing a formal field-theoretic representation of the partition function. Their findings also reproduced OS expression, while also accounting for the effects of ion-specific properties through the salt-exclusion layer thickness. In a separate study~\cite{dean2004field} by the same authors, they analyzed a system consisting of two interacting surfaces. This investigation specifically addressed the short-range interaction between a cation and the surface. Through this analysis, the researchers examined how the cation-specific interaction influenced both the effective charge and the disjoining pressure as a function of the separation between the surfaces. The study provided a detailed understanding of the implications of this specific interaction on the system's behavior.

We would like to emphasize that all models described above incorporate surface tension as surface free energy, which is calculated using thermodynamics and statistical physics of solutions. This approach can be referred to as a {\sl thermodynamic} or {\sl energetic} approach. However, according to the statistical theory of capillarity~\cite{rowlinson2013molecular,derjaguin1987derjaguin}, surface tension can also be derived from the mismatch between tangential and normal stress occurring in the adsorption layers. This approach can be tentatively labeled as the {\sl mechanical} approach. However, in order to employ this methodology, it is essential to have a technique capable of calculating the components of the stress tensor. One potential method for analyzing microscopic stress is the Irving-Kirkwood microscopic stress tensor method~\cite{irving1950statistical,shi2023perspective,rusanov2001condition,rusanov2001three}. It involves averaging of the microscopic stress tensor over the Gibbs distribution function, which is quite a difficult task in itself~\cite{shi2023perspective}. Recently, we developed another technique, presented in the papers~\cite{budkov2022modified,brandyshev2023noether,budkov2023macroscopic,vasileva2023theory,budkov2023dielectric,budkov2023variational,brandyshev2023statistical}, which involves deriving the stress tensor from the grand thermodynamic potential. This is done by viewing the latter as a functional of suitable order parameters and application of the Noether's theorems~\cite{hermann2022noether}. Since this approach self-consistently reunites the concepts of the grand thermodynamic potential and the stress tensor, it can be called a {\sl thermomechanical} approach.

Recently~\cite{budkov2023variational}, we utilized Wang's variational field theory~\cite{wang2010fluctuation} and discussed thermomechanical approach to derive the stress tensor from the grand thermodynamic potential of an inhomogeneous Coulomb fluid. Obtained stress tensor takes into account not only the osmotic pressure of the ions but also the electrostatic interactions beyond the mean-field approximation. In present study, using this stress tensor and considering the specific adsorption of ions onto the dielectric interface, we derive an analytical expression for the surface tension of electrolyte solutions in coexistence with nonpolar dielectric media. To validate our findings, we apply this expression to a large dataset of experimental data, obtaining very good approximations.

\section{Theoretical background}
Let us consider two phases as two continuous media with uniform dielectric permittivities: $\varepsilon_s$ for highly polar media (at $z>0$) and $\varepsilon_m$ for the non-polar media ($z<0$). There is a sharp planar boundary between two phases at $z=0$ ($z$-axis is perpendicular to the interface). In the highly polar media ($\varepsilon_s \gg 1$), there are dissolved ions with charges $q_{\alpha}$, where $\alpha$ enumerates different types of ions. The non-polar media with permittivity $\varepsilon_m \sim 1$ does not contain ions. We aim to utilize this model to elucidate the influence of the electrostatic interaction of dissolved ions and their adsorption onto the interface on the surface tension. This calculation will be performed using the stress tensor, derived recently~\cite{budkov2023variational} within variational field theory~\cite{wang2010fluctuation}
\begin{equation}
\label{Tij}
T_{ik}=-P\delta_{ik}+\varepsilon_s\left(\partial_i\psi\partial_k\psi-
\frac{1}{2}\delta_{ik}\partial_l\psi\partial_l\psi\right)+\varepsilon_s\left(\frac{1}{2}\mathcal{D}_{ll}(\bold{r})\delta_{ik}- \mathcal{D}_{ik}(\bold{r})\right),
\end{equation}
where $P$ is the local osmotic pressure of the ions, $\psi$ is the electrostatic potential, and the following tensor function
\begin{equation}
\mathcal{D}_{ik}(\bold{r})= k_{B}T\lim_{\bold{r}'\rightarrow \bold{r}}
\partial_i\partial_k' G(\bold{r},\bold{r}')
\end{equation}
has been introduced; $G(\bold{r},\bold{r}')$ is the trial Green function (see Appendix); $\partial_{i}=\partial/\partial x_{i}$ is the partial coordinate derivative. In eq. (\ref{Tij}) the first term in the right-hand side represents the hydrostatic isotropic stress tensor; the second term represents the standard Maxwell stress tensor; the third term represents the contribution of fluctuations in the local electric field around the mean-field configuration. Note that in eq. (\ref{Tij}) the summation over repeated indices is implied. Consequently, we determine the excess surface tension for a flat interface as follows~\cite{rowlinson2013molecular}
\begin{equation}
    \label{alpha}
    \Delta \alpha=\int\limits_0^{\infty}dz\,(P_{n}-P_{\tau})=\int\limits_0^{\infty}dz\,(T_{xx}-T_{zz}),
\end{equation}
where $P_{n}=-T_{zz}$ and $P_{\tau}=-T_{xx}=-T_{yy}$ are the local normal and tangential pressures, respectively, which can be determined via the components of the stress tensor (\ref{Tij})
Thus, we have
\begin{equation}
\label{T-T}
T_{xx}-T_{zz} = -\varepsilon_s\psi^{\prime}{}^2+
    \varepsilon_s\int\frac{d^2\mathbf{q}}{(2\pi)^2}\left(\mathcal{D}_{zz}(q,z)-\frac{1}{2}q^2Q(q,z)\right),
\end{equation}
where the first term in the right hand side stems from the Maxwell stress tensor, whereas the second one - from the fluctuation stress tensor; we have also introduced the following short-hand notations~\cite{budkov2023variational}
\begin{equation}
\label{Q}
Q(q,z)=k_{B}TG(q|z,z),
\end{equation}
\begin{equation}
\label{D}
\mathcal{D}_{zz}(q,z)=k_BT\lim_{z'\to z}\partial_z\partial_{z'}G(q|z,z').
\end{equation}

We consider the case of weak electrostatic interactions, so that the local electrostatic potential at $z>0$ can be obtained from the linearized PB equation
\begin{equation}
\label{linear_PB}
\psi^{\prime\prime}-\kappa^2\psi = 0
\end{equation}
with the following boundary conditions
\begin{equation}
\label{bound_cond}
\varepsilon_s\psi^{\prime}(0)=-\sigma_s,~\psi^{\prime}(\infty)=0,
\end{equation}
where 
\begin{equation}
\sigma_{s}=\sum\limits_{\alpha}q_{\alpha}n_{\alpha}b_{\alpha}e^{-\beta q_{\alpha}\psi_0}
\end{equation}
is the effective surface charge density of adsorbed ions and 
\begin{equation}
\label{kappa}
\kappa=\left(\frac{1}{\varepsilon_s k_{B}T}\sum\limits_{\alpha}q_{\alpha}^2 n_{\alpha}\right)^{1/2}
\end{equation}
is inverse Debye radius, $n_{\alpha}$ are the bulk ionic concentrations, and $b_{\alpha}$ represents the values with a dimension of length (see also Appendix), which is determined by the integral
\begin{equation}
b_{\alpha}=\int\limits_{0}^{\infty}dz\left(e^{-\beta u_{\alpha}(z)}-1\right)
\end{equation}
with $u_{\alpha}(z)$ being the effective potential of interaction between $\alpha$th ion and the dielectric boundary~\cite{shchekin2005thermodynamics,dos2010surface,dos2012ions} (or potential of mean force); parameters $b_{\alpha}$ determine the ionic specificity in this model and can be seen as measures of the adhesion of ions to the interface.

The solution of (\ref{linear_PB}) is
\begin{equation}
\label{psi}
\psi(z)=\psi_0 e^{-\kappa z},
\end{equation}
where the surface potential, $\psi_0$, is determined by the first boundary condition (\ref{bound_cond}), resulting in the following transcendental equation
\begin{equation}
\psi_0=\frac{1}{\varepsilon_s \kappa}\sum\limits_{\alpha}q_{\alpha}n_{\alpha}b_{\alpha}e^{-\beta q_{\alpha}\psi_0}.
\end{equation}
In the linear approximation (where $\beta q_{\alpha}\psi_0 \ll 1$) the latter yields 
\begin{equation}
\label{psi0}
\psi_0=\frac{\frac{1}{\varepsilon_s \kappa}\sum\limits_{\alpha}q_{\alpha}n_{\alpha}b_{\alpha}}{1+\frac{1}{\varepsilon_s \kappa k_{B}T}\sum\limits_{\alpha}q_{\alpha}^2n_{\alpha}b_{\alpha}}.
\end{equation}
At $z<0$ the Green function is 
\begin{equation}
G(q|z,z')=A e^{qz}.
\end{equation}
At $z>0$ the Green function satisfies the following equation (see also eq. (\ref{eq_for_G}) in Appendix)
\begin{equation}
    \label{Green_func_eq}
    \varepsilon_{s}(-\partial_z^2+\varkappa^2_{q}+\kappa_s\delta(z))G(q|z,z')=\delta(z-z'),
\end{equation}
where $\varkappa_{q}=\sqrt{\kappa^2+q^2}$ and
\begin{equation}
\label{kappa_s}
\kappa_s=\frac{1}{\varepsilon_s}\sum\limits_{\alpha}q_{\alpha}^2n_{\alpha}b_{\alpha}\left(1-\beta q_{\alpha}\psi_0\right).
\end{equation}
The solution of eq. (\ref{Green_func_eq}) is
\begin{equation}
    G(q|z,z')=
    \begin{cases}
        B\sinh{\varkappa_qz}+C\cosh{\varkappa_qz},& 0<z<z',\\
        D\,e^{-\varkappa_qz},& z>z'.      
    \end{cases}
\end{equation}
The coefficients $A$, $B$, $C$, $D$ have to be obtained from the boundary conditions:
\begin{align}
G(q|-0,z')&-G(q|+0,z')=0,\\
\varepsilon_m\frac{\partial G(q|-0,z')}{\partial z}&-\varepsilon_s\frac{\partial G(q|+0,z')}{\partial z}+\varepsilon_s\kappa_s G(q|0,z')=0,\\
G(q|z'-0,z')&-G(q|z'+0,z')=0,\\
\varepsilon_s\bigg{(}\frac{\partial G(q|z'-0,z')}{\partial z}&-\frac{\partial G(q|z'+0,z')}{\partial z}\bigg{)}=1.
\end{align}
Thus, the solution is
\begin{equation}
G(q|z,z')=
    \begin{cases}
        \frac{\varepsilon_m e^{-\varkappa_qz'}}{\varepsilon_m q+\varepsilon_s(\varkappa_q+\kappa_s))}e^{qz},& z<0,\\
       \frac{e^{-\varkappa_q z'}\left((\varepsilon_m q+\varepsilon_s \kappa_s)\sinh{\varkappa_q z}+\varepsilon_s\varkappa_q \cosh{\varkappa_q z}\right)}{\varepsilon_s\varkappa_q\left(\varepsilon_m q+\varepsilon_s(\kappa_s+\varkappa_q)\right)},& 0<z<z',\\
        \frac{e^{-\varkappa_q z}\left((\varepsilon_m q+\varepsilon_s \kappa_s)\sinh{\varkappa_q z'}+\varepsilon_s\varkappa_q \cosh{\varkappa_q z'}\right)}{\varepsilon_s\varkappa_q\left(\varepsilon_m q+\varepsilon_s(\kappa_s+\varkappa_q)\right)},& z>z'>0.      
    \end{cases}
\end{equation}

For further calculations of the surface tension, we have to subtract from the Green function the Green function of infinite media $G({q|z,z'})=e^{-\varkappa_q |z-z'|}/{2\varepsilon_s\varkappa_q}$ in the region of $z>0$, that yields
\begin{equation}
G(q|z,z')=\frac{e^{-\varkappa_q(z+z')}\left(\varepsilon_s(\varkappa_q-\kappa_s)-\varepsilon_m q\right)}{2\varepsilon_s\varkappa_q\left(\varepsilon_m q+\varepsilon_s(\kappa_s+\varkappa_q)\right)}.   
\end{equation}
Therefore, we have 
\begin{equation}
\label{Q_2}
Q(q,z)=k_{B}T\frac{\left(\varepsilon_s(\varkappa_q-\kappa_s)-\varepsilon_m q\right)}{2\varepsilon_s\varkappa_q\left(\varepsilon_m q+\varepsilon_s(\kappa_s+\varkappa_q)\right)}e^{-2\varkappa_qz},
\end{equation}
\begin{equation}
\label{D_2}
\mathcal{D}_{zz}(q,z)=k_{B}T\frac{\varkappa_q\left(\varepsilon_s(\varkappa_q-\kappa_s)-\varepsilon_m q\right)}{2\varepsilon_s\left(\varepsilon_m q+\varepsilon_s(\kappa_s+\varkappa_q)\right)}e^{-2\varkappa_qz}.
\end{equation}

Thus, substituting (\ref{Q_2}), (\ref{D_2}), and (\ref{psi}) into (\ref{T-T}) and then integrating over $z$ in (\ref{alpha}), we eventually obtain
\begin{equation}
\label{alpha2}
\Delta\alpha=-\frac{\varepsilon_s \kappa \psi_0^2}{2}+\frac{k_{B}T}{8}\int\frac{d^2\bold{q}}{(2\pi)^2} \frac{\left(\varepsilon_s(\sqrt{\kappa^2+q^2}-\kappa_s)-\varepsilon_m q\right)(2\kappa^2+q^2)}{(\kappa^2+q^2)\left(\varepsilon_m q+\varepsilon_s(\kappa_s+\sqrt{\kappa^2+q^2})\right)}.
\end{equation}
In the absence of ions ($\psi_0=0$, $\kappa_s=\kappa=0$) eq. (\ref{alpha2}) transforms into
\begin{equation}
\Delta\alpha^{(0)}=\frac{k_{B}T}{8}\int\frac{d^2\bold{q}}{(2\pi)^2} \frac{\varepsilon_s-\varepsilon_m}{\varepsilon_s+\varepsilon_m}
\end{equation}
that is nothing but a contribution of the classical van der Waals interactions (or zeroth Matsubara mode)~\cite{dzyaloshinskii1961general,ninham1997ion,hatlo2008role} to the surface tension which have to be subtracted from the final result. Therefore, we get
\begin{multline}
\label{alpha_fin}
\Delta\alpha=-\frac{\varepsilon_s \kappa \psi_0^2}{2}+\\\frac{k_{B}T}{16\pi}\frac{\varepsilon_s-\varepsilon_m}{\varepsilon_s+\varepsilon_m}\int\limits_{0}^{\Lambda}dq \,q \left(\frac{(\varepsilon_s+\varepsilon_m)\left(\varepsilon_s(\sqrt{\kappa^2+q^2}-\kappa_s)-\varepsilon_m q\right)(2\kappa^2+q^2)}{(\varepsilon_s-\varepsilon_m)(\kappa^2+q^2)\left(\varepsilon_m q+\varepsilon_s(\kappa_s+\sqrt{\kappa^2+q^2})\right)}-1\right),
\end{multline}
where we have introduced the untraviolet cut-off parameter, $\Lambda$, which is related to the atomic scale length and will be specified below. Introducing a new variable of integration, $x=q/\kappa$, in eq. (\ref{alpha_fin}), we can obtain
\begin{equation}
\label{delta_alpha}
\Delta\alpha=-\frac{\varepsilon_s \kappa \psi_0^2}{2}+\frac{k_{B}T\kappa^2}{16\pi}\frac{\varepsilon_s-\varepsilon_m}{\varepsilon_s+\varepsilon_m}f(\tilde{\Lambda},\delta,\varepsilon_{sm}),
\end{equation}
where we have introduced the auxiliary function
\begin{equation}
\label{f}
f(\tilde{\Lambda},\delta,\varepsilon_{sm})=\int\limits_{0}^{\tilde{\Lambda}}dx\, x \left(\frac{(\varepsilon_{sm}+1)\left(\varepsilon_{sm}(\sqrt{1+x^2}-\delta)-x\right)(2+x^2)}{(\varepsilon_{sm}-1)(1+x^2)\left(x+\varepsilon_{sm}(\delta+\sqrt{1+x^2})\right)}-1\right),
\end{equation}
with the arguments $\tilde{\Lambda}=\Lambda/\kappa$, $\delta=\kappa_s/\kappa$, and $\varepsilon_{sm}=\varepsilon_{s}/\varepsilon_{m}$. Eqs. (\ref{psi0}), (\ref{delta_alpha}), and (\ref{f}) provide the solution to the posed problem within the thermomechanical approach.

In the absence of adsorbed ions ($\kappa_s=0$, $\psi_0=0$), we have 
\begin{equation}
\label{alpha_OS}
\Delta\alpha=\frac{k_{B}T\kappa^2}{16\pi}\frac{\varepsilon_s-\varepsilon_m}{\varepsilon_s+\varepsilon_m}g(\tilde{\Lambda},\varepsilon_{sm}),
\end{equation}
where $g(\tilde{\Lambda},\varepsilon_{sm})=f(\tilde{\Lambda},0,\varepsilon_{sm})$. At sufficiently small concentration, when $\tilde{\Lambda}\gg 1$ the function $g(\tilde{\Lambda},\varepsilon_{sm})$ has the following asymptotic 
\begin{equation}
  g(\tilde{\Lambda},\varepsilon_{sm})\simeq (1+O(\varepsilon_{m}/\varepsilon_{s}))\ln\tilde{\Lambda}  
\end{equation}
that leads to a well-known Onsager-Samaras result~\cite{onsager1934surface,shchekin2005thermodynamics,levin2009ions,markovich2014surface}
\begin{equation}
\Delta\alpha_{OS}\simeq \frac{k_{B}T\kappa^2}{16\pi}\frac{\varepsilon_s-\varepsilon_m}{\varepsilon_s+\varepsilon_m}\ln\tilde{\Lambda}.
\end{equation}

{To conclude this section, let us clarify the physical interpretation of the Green function $G(q | z, z')$ in our particular problem. From the expression (\ref{n_alpha}) in the Appendix (see also \cite{budkov2023variational}), we can see that the average ionic concentrations are given by: 
\begin{equation}
\bar{n}_\alpha(z) = n_\alpha e^{-\beta q_\alpha \psi(z)-\frac{\beta q^2_\alpha G(0 ; z,z')}{2}},
\end{equation}
where the function
\begin{equation}
\label{G_image}
G(0;z,z)=\int \frac{d^2\bold{q}}{(2\pi)^2}\frac{e^{-2\varkappa_qz }\left(2\varepsilon_s(\varkappa_q-\kappa_s)-\varepsilon_m q\right)}{\varepsilon_s\varkappa_q\left(\varepsilon_m q+\varepsilon_s(\kappa_s+\varkappa_q)\right)}
\end{equation}
is determined by considering the subtracted Green function for the infinite medium, and we take into account that 
\begin{equation}
\label{n_alphabulk}
n_{\alpha}=z_{\alpha}\exp\left[-\frac{\beta q_{\alpha}^2}{4\varepsilon_s}\int\frac{d^2\bold{q}}{(2\pi)^2}\left(\frac{1}{\kappa_q}-\frac{1}{q}\right)\right]= z_{\alpha}\exp\left[\frac{\beta q_{\alpha}^2\kappa}{8\pi\varepsilon_s}\right],
\end{equation}
and that
\begin{equation}
G_{0}(0)=\frac{1}{2\varepsilon_s}\int \frac{d^{2}\bold{q}}{(2\pi)^2}\frac{1}{q}.
\end{equation}
From (\ref{n_alphabulk}) we have obtained expressions for the bulk chemical potentials in Debye-H{\"u}ckel theory approximation
\begin{equation}
\mu_{\alpha}=k_{B}T\ln(n_{\alpha}\Lambda_{\alpha}^3)-\frac{q_{\alpha}^2\kappa}{8\pi\varepsilon_s}.
\end{equation}
In the absence of adsorbed ions and when $\kappa z \ll 1$, we obtain 
\begin{equation}
G(0; z, z) = \frac{1}{8\pi \varepsilon_s} \frac{\varepsilon_s - \varepsilon_m}{\varepsilon_s + \varepsilon_m} \frac{1}{z}.
\end{equation}
This equation determines nothing more than the potential energy associated with image forces~\cite{landau2013electrodynamics}, i.e.
\begin{equation}
u_{\alpha}^{(im)}(z)=\frac{q_{\alpha}^2}{2}G(0;z,z)=\frac{q_{\alpha}^2}{16\pi\varepsilon_s z}\frac{\varepsilon_s-\varepsilon_m}{\varepsilon_s+\varepsilon_m}
\end{equation}
At $\varepsilon_s\gg \varepsilon_c$, the integral (\ref{G_image}) yields
\begin{equation}
G(0;z,z)=\frac{1}{16\pi\varepsilon_s}\frac{e^{-2\kappa z}}{z},
\end{equation}
that in turn gives the following image force potential with account for electrostatic screening
\begin{equation}
u_{\alpha}^{(im)}(z)=\frac{q_{\alpha}^2}{2}G(0;z,z)=\frac{q_{\alpha}^2}{16\pi\varepsilon_s z}e^{-2\kappa z}.
\end{equation}
The heuristic derivation of this expression was first proposed by Onsager and Samaras in ref.~\cite{onsager1934surface}, and it was obtained within the framework of statistical field theory in refs.~\cite{wang2016inhomogeneous,Lau2020,hatlo2008role}.}

{Thereby, the function
\begin{equation}
u_{\alpha}^{(im)}(z)=\frac{q_{\alpha}^2}{2}G(0;z,z)=\frac{ q_{\alpha}^2}{2\varepsilon_s}\int \frac{d^2\bold{q}}{(2\pi)^2}\frac{\left(\varepsilon_s(\varkappa_q-\kappa_s)-\varepsilon_m q\right)e^{-2\varkappa_qz }}{\varkappa_q\left(\varepsilon_m q+\varepsilon_s(\kappa_s+\varkappa_q)\right)}
\end{equation}
can be considered as the potential energy of image forces with account for electrostatic screening and adsorption of the ions on the dielectric boundary within the variational field theory.}

\section{Application to aqueous 1:1 electrolyte solutions}
Let us consider a case of the one-component aqueous 1:1 electrolyte solution coexisting with nonpolar dielectric media. In this case, $q_{\pm}=\pm e$ ($e$ is the elementary charge). Following the papers~\cite{shchekin2005thermodynamics,markovich2014surface,markovich2015surface}, we assume that only anions adsorb onto the interface. In this case $b_{+}=0$ and $b_{-}=-b$. Thus, eq. (\ref{psi0}) transforms into
\begin{equation}
\psi_0=\frac{enb}{\varepsilon_s\kappa \left(1-\frac{\kappa b}{2}\right)},
\end{equation}
where $\kappa = (2e^2n/k_{B}T\varepsilon_s)^{1/2}$ and $n=n_{\pm}$ is the bulk concentration of the ions. As it follows from (\ref{psi0}) and (\ref{kappa_s}), the parameter $\delta=\kappa_s/\kappa$ is determined by
\begin{equation}
\delta = -\frac{\kappa b}{4\left(1-\frac{\kappa b}{2}\right)}.
\end{equation}

Based on the papers by Markovich et al.~\cite{markovich2014surface,markovich2015surface,markovich2017surface}, we adopt the relation for the ultraviolet cut-off parameter $\Lambda=2\sqrt{\pi}/a$. Here, $a$ represents the average minimal distance of approach between ions, determined by the summation of the hydration radii, $r_{\pm}$. The values for $r_{\pm}$ are obtained from~\cite{nightingale1959phenomenological}. Then, using $b$ as an adjustable parameter, we fit a number of available in literature experimental data with the help of eq. (\ref{delta_alpha}).

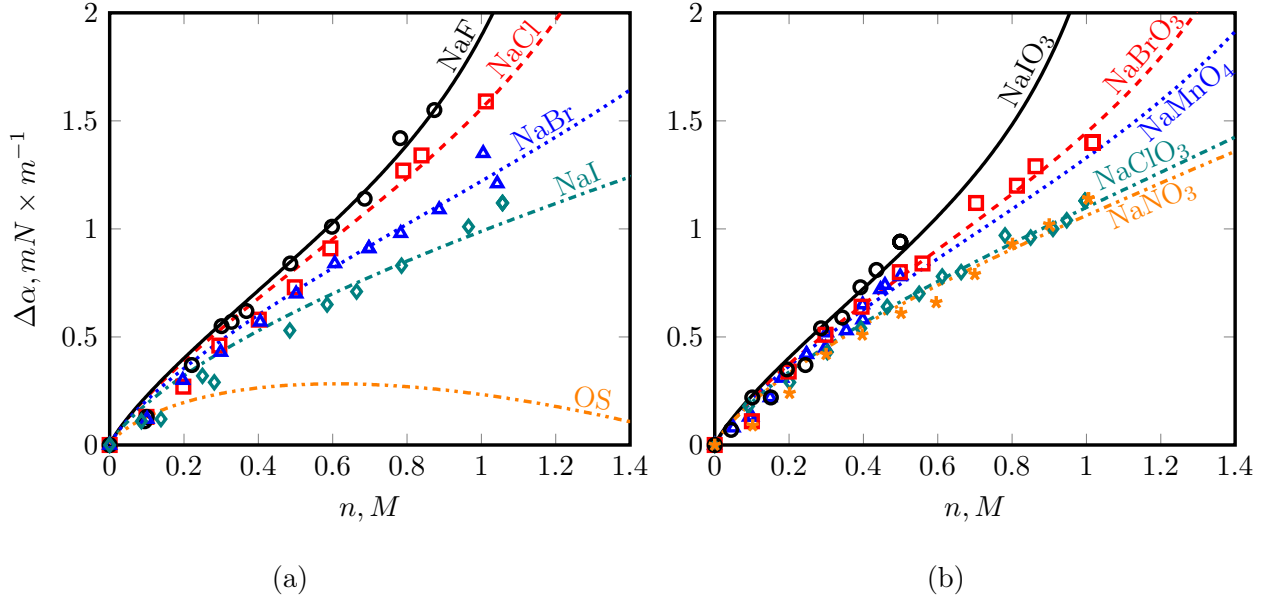
\begin{figure}[ht]
\begin{subfigure}[b]{0.47\textwidth}
\begin{tikzpicture}
        \begin{axis}[
            xlabel={$n, M$},
            ylabel={$\Delta\alpha,mN\times m^{-1}$},
            xmin=0,
            xmax=1.4,
            ymin=0,
            ymax=2,
            width=8.5cm,
            ]
        \addplot [only marks,mark=square,draw=red] table [x=x1,y=NaCl] {pics_data/Na.dat};
        \addplot [only marks,mark=o,draw=black] table [x=x2,y=NaF] {pics_data/Na.dat};
        \addplot [only marks,mark=triangle,draw=blue] table [x=x3,y=NaBr] {pics_data/Na.dat};
        \addplot [only marks,mark=diamond,draw=teal] table [x=x4,y=NaI] {pics_data/Na.dat};
        \addplot [red,dashed] table [x=x,y=NaCl] {pics_data/Na_fit.dat}
        node [pos=.62,yshift=6pt,sloped] {NaCl};
        \addplot [black,solid] table [x=x,y=NaF] {pics_data/Na_fit.dat}
        node [pos=.22,yshift=6pt,sloped] {NaF};
        \addplot [blue,dotted] table [x=x,y=NaBr] {pics_data/Na_fit.dat}
        node [pos=.8,yshift=6pt,sloped] {NaBr};
        \addplot [teal,dashdotted] table [x=x,y=NaI] {pics_data/Na_fit.dat}
        node [pos=.87,yshift=6pt,sloped] {NaI};
        \addplot [orange,dashdotdotted] table [x=x,y=NaOS] {pics_data/Na_fit.dat}
        node [pos=.87,yshift=6pt,sloped] {OS};
        \end{axis}
\end{tikzpicture}
\caption{}
\label{sodium+anions_1}
\end{subfigure}
\hfill
\begin{subfigure}[b]{0.47\textwidth}
\begin{tikzpicture}
    \begin{axis}[
        xlabel={$n, M$},
        xmin=0,
        xmax=1.4,
        ymin=0,
        ymax=2,
        width=8.5cm,
        ]
    \addplot [only marks,mark=triangle,draw=blue] table [x=x1,y=NaMnO4] {pics_data/sodium_salts.dat};
    \addplot [only marks,mark=diamond,draw=teal] table [x=x2,y=NaClO3] {pics_data/sodium_salts.dat};
    \addplot [only marks,mark=square,draw=red] table [x=x3,y=NaBrO3] {pics_data/sodium_salts.dat};
    \addplot [only marks,mark=o,draw=black] table [x=x4,y=NaIO3] {pics_data/sodium_salts.dat};
    \addplot [only marks,mark=star,draw=orange] table [x=x5,y=NaNO3] {pics_data/Na.dat};
    \addplot [blue,dotted] table [x=x,y=NaMnO4] {pics_data/sodium_salts_fit.dat}
    node [pos=0.8,yshift=-6pt,sloped] {NaMnO$_4$};
    \addplot [teal,dashdotted] table [x=x,y=NaClO3] {pics_data/sodium_salts_fit.dat}
    node [pos=0.8,yshift=6pt,sloped] {NaClO$_3$};
    \addplot [red,dashed] table [x=x,y=NaBrO3] {pics_data/sodium_salts_fit.dat}
    node [pos=0.7,yshift=6pt,sloped] {NaBrO$_3$};
    \addplot [black,solid] table [x=x,y=NaIO3] {pics_data/sodium_salts_fit.dat}
    node [pos=0.25,yshift=6pt,sloped] {NaIO$_3$};
    \addplot [orange,dashdotdotted] table [x=x,y=NaNO3] {pics_data/sodium_salts_fit.dat}
    node [pos=0.8,yshift=-6pt,sloped] {NaNO$_3$};
    \end{axis}
\end{tikzpicture}
\caption{}
\label{sodium+anions_2}
\end{subfigure}
\caption{Fit of experimental \cite{matubayasi2013surface} excess surface tension of electrolyte solutions containing sodium cation, T=298 K, $\varepsilon_s=78\varepsilon_0$, $\varepsilon_m=1\varepsilon_0$ ($\varepsilon_0$ is the vacuum permittivity).}
\label{sodium_anions}
\end{figure}

\begin{figure}[ht]
\begin{subfigure}[b]{0.47\textwidth}
\begin{tikzpicture}
        \begin{axis}[
            xlabel={$n, M$},
            ylabel={$\Delta\alpha,mN\times m^{-1}$},
            xmin=0,
            xmax=1.4,
            ymin=0,
            ymax=2,
            width=8.5cm,
            ]
         \addplot [black,solid] table [x=x,y=KCl] {pics_data/potassium_salts_fit.dat} node [pos=0.65,yshift=6pt,sloped] {KCl};
    \addplot [blue,dotted] table [x=x,y=KNO3] {pics_data/potassium_salts_fit.dat} node [pos=0.8,yshift=6pt,sloped] {KNO$_3$};
    \addplot [teal,dashdotted] table [x=x,y=KI] {pics_data/potassium_salts_fit.dat} node [pos=0.8,yshift=-6pt,sloped] {KI};
    \addplot [red,dashed] table [x=x,y=KBr] {pics_data/potassium_salts_fit.dat} node [pos=0.8,yshift=6pt,sloped] {KBr};
    \addplot [orange,dashdotdotted] table [x=x,y=KMnO4] {pics_data/potassium_salts_fit.dat} node [pos=0.77,yshift=6pt,sloped] {KMnO$_4$};
    \addplot [only marks,mark=o,draw=black] table [x=x1,y=KCl] {pics_data/potassium_salts.dat};
    \addplot [only marks,mark=triangle,draw=blue] table [x=x2,y=KNO3] {pics_data/potassium_salts.dat};
    \addplot [only marks,mark=diamond,draw=teal] table [x=x3,y=KI] {pics_data/potassium_salts.dat};
    \addplot [only marks,mark=square,draw=red] table [x=x4,y=KBr] {pics_data/potassium_salts.dat};
    \addplot [only marks,mark=star,draw=orange] table [x=x5,y=KMnO4] {pics_data/potassium_salts.dat};
        \end{axis}
\end{tikzpicture}
\caption{ }
\label{potatssium+anions_m}
\end{subfigure}
\hfill
\begin{subfigure}[b]{0.47\textwidth}
\begin{tikzpicture}
    \begin{axis}[
        xlabel={$n, M$},
        xmin=0,
        xmax=1.4,
        ymin=0,
        ymax=2.5,
        width=8.5cm,   
        ]
    \addplot [only marks,mark=o,draw=black] table [x=x1,y=KCl] {pics_data/K.dat};
    \addplot [only marks,mark=triangle,draw=blue] table [x=x2,y=KI] {pics_data/K.dat};
    \addplot [only marks,mark=square,draw=red] table [x=x3,y=KBr] {pics_data/K.dat};
    \addplot [black,solid] table [x=x,y=KCl] {pics_data/K_fit.dat}
    node [pos=.75,yshift=6pt,sloped] {KCl};
    \addplot [blue,dotted] table [x=x,y=KI] {pics_data/K_fit.dat}
    node [pos=.85,yshift=6pt,sloped] {KI};
    \addplot [red,dashed] table [x=x,y=KBr] {pics_data/K_fit.dat}
    node [pos=.8,yshift=-6pt,sloped] {KBr};
    \end{axis}
\end{tikzpicture}
\caption{}
\label{potassium+anions_a}
\end{subfigure}
\caption{Fit of experimental \cite{matubayasi2013surface,aveyard1976interfacial} excess surface tension of electrolyte solutions containing potassium cation, T=298 K (\ref{potatssium+anions_m}) and 293 K (\ref{potassium+anions_a}), $\varepsilon_s=78\varepsilon_0$, $\varepsilon_m=1\varepsilon_0$.}
\label{potassium_anions}
\end{figure}
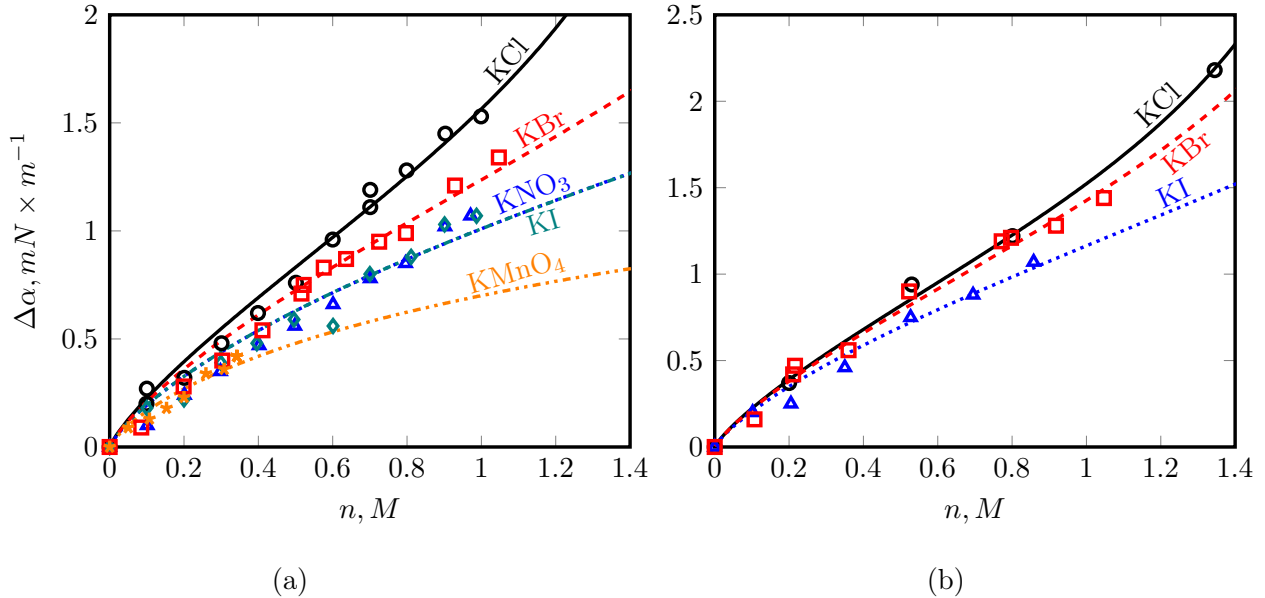

In Figs. \ref{sodium_anions} and \ref{potassium_anions} we present a comparison of calculated excess surface tension values with experimental results for a set of electrolyte solutions with Na$^+$ and K$^+$ cations at water-air interface, and also plot a well-known result of the OS theory (see Fig. \ref{sodium+anions_1}). Overall, we can conclude that the model successfully describes available data, with the exception of the region of rather small concentration. This, as we assume, can be related to the difficulties of experimental measurements in this range. Moreover, the experimental data from different sources are rather scattered even across the entire range of concentrations (refer, for instance, to the visual comparison at Figs. \ref{potassium_anions}, and check the fitting parameter values in Tables \ref{sodium_potassium} and \ref{w-a_w-ddcn}, where we fit the experiments from Refs. \cite{matubayasi2013surface} and \cite{aveyard1976interfacial}, correspondingly).

\begin{table}
\caption{Parameters of fitting for sodium and potassium containing aqueous solutions; experimental data were taken from \cite{matubayasi2013surface}; T=298 K, $\varepsilon_s=78\varepsilon_0$, $\varepsilon_m=1\varepsilon_0$.}
    \centering
    \begin{tabular}{|c|c|c|c|c|c|}
        salt    &$a\times10^{10}$, m &$b/a$&salt    &$a\times10^{10}$, m &$b/a$\\
        \hline
        NaCl	&6.9	&0.397    &KCl	    &6.62	&0.397\\
        NaF	    &7.1	&0.438&&&\\
        NaBr	&6.88	&0.319    &KBr	    &6.61	&0.315\\
        NaI	    &6.89	&0.236    &KI	    &6.62	&0.232\\
        NaNO$_3$	&6.93	&0.267    &KNO$_3$	    &6.66	&0.234\\
        NaMnO$_4$	&7.03	&0.352    &KMnO$_4$	&6.76	&0.081\\
        NaClO$_3$	&6.99	&0.283&&&\\
        NaBrO$_3$	&7.09	&0.378&&&\\
        NaIO$_3$	&7.32	&0.454&&&\\
    \end{tabular}
    \label{sodium_potassium}
\end{table}

\begin{figure}[ht]
\begin{tikzpicture}
\begin{axis}[
    xlabel={$n, M$},
    ylabel={$\Delta\alpha,mN\times m^{-1}$},
    xmin=0,
    xmax=1.4,
    ymin=-0.5,
    ymax=2.2,
    ]
\addplot [only marks,mark=o,draw=black] table [x=x2,y=KCl] {pics_data/w-ddcn.dat};
\addplot [only marks,mark=triangle,draw=blue] table [x=x3,y=KI] {pics_data/w-ddcn.dat};
\addplot [only marks,mark=square,draw=red] table [x=x4,y=KBr] {pics_data/w-ddcn.dat};

\addplot [black,solid] table [x=x,y=KCl] {pics_data/w-ddcn_fit.dat}
node [pos=.8,yshift=6pt,sloped] {KCl};
\addplot [blue,dotted] table [x=x,y=KI] {pics_data/w-ddcn_fit.dat}
node [pos=.8,yshift=6pt,sloped] {KI};
\addplot [red,dashed] table [x=x,y=KBr] {pics_data/w-ddcn_fit.dat}
node [pos=.85,yshift=6pt,sloped] {KBr};
\end{axis}
\end{tikzpicture}
\caption{Fit of the experimental \cite{aveyard1976interfacial} excess surface tension of electrolyte solutions with potassium cation at the water-dodecane interface, T=293 K, $\varepsilon_s=78\varepsilon_0$, $\varepsilon_m=2.01\varepsilon_0$.}
\label{aveyard_w-ddcn}
\end{figure}
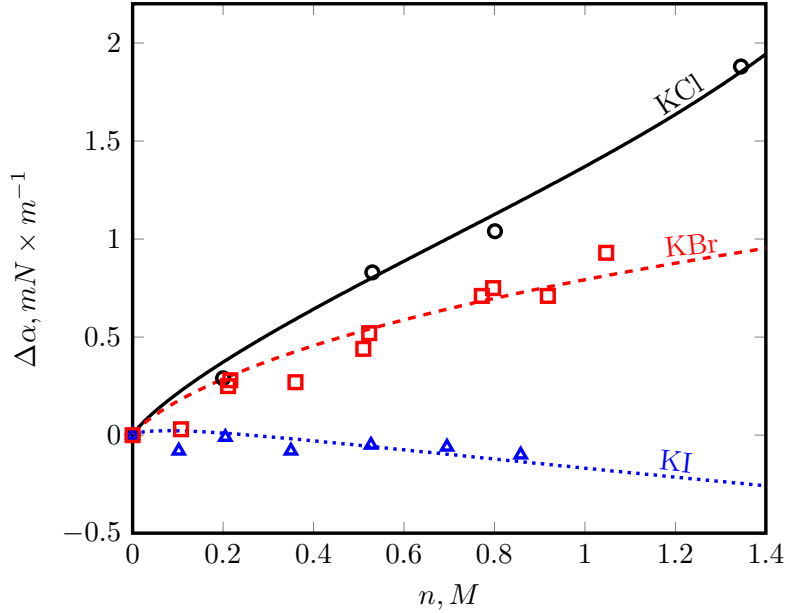

As is seen, we also can accurately describe the behavior of the surface tension of 1:1 electrolyte solutions at the oil-water interface (see Fig. \ref{aveyard_w-ddcn} and Table \ref{w-a_w-ddcn}). Moreover, we successfully fitted the surface tension of a number of 1:1 electrolytes, data for which were taken from Ref. \cite{matubayasi2013surface}. We do not present the graphs, but all parameters of calculations are presented in the Table \ref{remaining_stuff}.

\begin{table}
\caption{Fitting parameters for a set of electrolyte solutions at water-air and water-dodecane interfaces; experimental data were taken from \cite{aveyard1976interfacial}; T=293 K, $\varepsilon_s=78\varepsilon_0$, $\varepsilon_m=1\varepsilon_0$ for air and $\varepsilon_m=2.01\varepsilon_0$ - for dodecane.}
    \centering
    \begin{tabular}{|c|c|c|c|}
        salt    &$a\times10^{10}$,m &$b/a$, water-air &$b/a$, water-dodecane\\
        \hline
        NaCl	&6.9	&0.422	&0.361\\
        KCl	    &6.62	&0.387	&0.358\\
        KI	    &6.62	&0.292	&-1.037\\
        KBr	    &6.61	&0.366	&0.131\\
        LiCl	&7.14	&0.415	&0.363\\
    \end{tabular}
    \label{w-a_w-ddcn}
\end{table}

\begin{table}
\caption{Fitting parameters for a set of different systems, experimental data for which were taken from \cite{matubayasi2013surface}; T=298 K, $\varepsilon_s=78\varepsilon_0$, $\varepsilon_m=1\varepsilon_0$.}
    \centering
    \begin{tabular}{|c|c|c|}
        salt    &$a\times10^{10}$,m &$b/a$\\
        \hline
        LiCl	&7.14	&0.373\\
        LiNO$_3$   &7.17	&0.279\\
        NH$_4$NO   &6.66	&0.166\\
        NH$_4$Cl   &6.63	&0.292\\
        NH$_4$Br   &6.61	&0.273\\
        NH$_4$I	&6.62	&-0.083\\
        CsCl	&6.61	&0.408\\
        CuSO$_4$   &7.98	&0.429\\
        MgSO$_4$   &8.07	&0.413\\
        NiSO$_4$   &7.83	&0.425\\
    \end{tabular}
    \label{remaining_stuff}
\end{table}

\section{Discussion}

{First, we would like to note that, similar to previous studies \cite{markovich2014surface, dos2012ions}, we obtained not only the "standard" reverse Hofmeister series for anions at the water/air interface (F$^{-}$ > Cl$^{-}$ > Br$^{-}$ > I$^{-}$), but also discovered an "extended" series for the affinity parameter $b$. This extended series is as follows: IO$_{3}^{-}$ > F$^{-}$ > Cl$^{-}$ > BrO$_{3}^{-}$ > Br$^{-}$ > ClO$_{3}^{-}$ > NO$_{3}^{-}$ > I$^{-}$. This series slightly differs from both the one obtained by Markovich et al. \cite{markovich2015surface} and the one discovered by dos Santos et al. \cite{dos2010surface}, suggesting that the correct interpretation is still under question. This is especially true considering the limited experimentally investigated concentration range of some systems (NaIO$_{3}$, NaMnO$_{4}$), which could lead to ambiguity in the fitting process. The discrepancy between the previously presented data and the inverse Hofmeister series demonstrated above necessitates further detailed investigations into the potential of mean force of the anions at the water-air interface. This requires performing advanced full atomic computer simulations, including {\sl ab initio} calculations.}

{Second, we would like to discuss potential future developments of the current theory, highlighting its flexibility and ease of use. The derived formulas (\ref{psi0}), (\ref{delta_alpha}) and (\ref{f}), can be conveniently applied to assess surface tension of multicomponent electrolyte solutions, making them a valuable tool in evaluating surface tension within these systems. Moreover, the application of the theory can be expanded beyond weak electrostatic coupling. By solving the self-consistent field equations numerically, as demonstrated in recent publications~\cite{wang2015theoretical,wang2016inhomogeneous}, it becomes possible to study electrolyte systems containing divalent ions, including those with complex electric structures~\cite{urbanija2008attraction} and room-temperature ionic liquids~\cite{kondrat2023theory,budkov2022electric}. However, in the strong coupling regime, it is essential to take into consideration the effect of dielectric saturation in the solvent near the charged interface, as discussed in literature~\cite{gongadze2012decrease,gongadze2014ions,marcovitz2015water,budkov2018theory,dubtsov2018liquid}. It could be incorporated into an explicit solvent model, similar to those proposed in papers~\cite{gongadze2012decrease,gongadze2014ions,abrashkin2007dipolar,budkov2022modified}, but going beyond the mean field approximation. On the other hand, theoretical description of the room-temperature ionic liquids is even more complex, requiring consideration of excluded volume and structural effects relating to the molecular structure of ions\cite{kondrat2023theory,budkov2022electric,cruz2019effect,cruz2021phase,vasileva2023theory,blossey2017structural}. This opens up new research avenues and could be the topic of future publications.}

\section{Conclusion}
In conclusion, we have utilized the thermomechanical approach to derive the surface tension of aqueous electrolyte solutions in contact with non-polar dielectric media. The surface tension has been obtained by calculating both the normal and tangential pressure using the stress tensor components derived within the variational field theory. Using this approach, in the linear approximation, taking into account specific adsorption of anions onto the dielectric boundary we have derived an analytical expression for surface tension. We have demonstrated that at low electrolyte concentrations, the derived expression follows the classical Onsager-Samaras limiting law. We have fitted various experimental data related to the surface tension of the solution-air and solution-dodecane interface by utilizing only one fitting parameter, which corresponds to the affinity of anions to the dielectric interface. 

\textbf{Data availability statement.} {\sl The data supporting the findings of this study are available in the article.}

{\bf Acknowledgements.}  
This work is an output of a research project implemented as part of the Basic Research Program at the National Research University Higher School of Economics (HSE University). The numerical calculations were partially performed on the supercomputer facilities provided by NRU HSE.

\appendix
\section{Derivation of basic equations}
As it follows from the variational field theory~\cite{wang2010fluctuation,budkov2023variational}, the grand thermodynamic potential, $\Omega$, of the electrolyte solution within the model of Coulomb gas can be approximated by the following functional
\begin{multline}
-\frac{\Omega}{k_{B}T}=W[G;\psi]=\frac{1}{2}\ln\frac{Det G}{Det G_{0}}-\frac{1}{2}tr\left(G[G_0^{-1}-G^{-1}]\right)+\frac{\beta}{2}(\psi G_0^{-1}\psi)+\\\sum\limits_{\alpha}z_{\alpha}\int d\bold{r} \theta(z)A_{\alpha}(z)e^{-\beta q_{\alpha}\psi(\bold{r}) - \beta u_{\alpha}(z)}=\frac{1}{2}\ln\frac{Det G}{Det G_{0}}-\frac{1}{2}tr\left(G[G_0^{-1}-G^{-1}]\right)+\frac{\beta}{2}(\psi G_0^{-1}\psi)+\\\sum\limits_{\alpha}z_{\alpha}\int d\bold{r}\theta(z) A_{\alpha}(z)e^{-\beta q_{\alpha}\psi(\bold{r})}+\sum\limits_{\alpha}z_{\alpha}\int d\bold{r}\theta(z)A_{\alpha}(z)e^{-\beta q_{\alpha}\psi(\bold{r})}(e^{-\beta u_{\alpha}(z)}-1),
\end{multline}
where $\beta = (k_{B}T)^{-1}$ and $z_{\alpha}=\Lambda_{\alpha}^{-3}e^{\beta\mu_{\alpha}}\theta_{\alpha}$ are the ionic fugacities expressed via the chemical potentials, $\mu_{\alpha}$, the de
Broglie thermal wavelengths, $\Lambda_{\alpha}$, and internal partition functions, $\theta_{\alpha}$, of the ions; $\theta(z)$ is the standard Heaviside step function. $Det G$ and $Det G_0$ are the functional determinants of the corresponding Green functions and $tr(...)$ is the trace of operator~\cite{budkov2023variational}. We have also introduced the following short-hand notation
\begin{equation}
(\psi G_0^{-1}\psi)=\int d\bold{r}\int d\bold{r}^{\prime} \psi(\bold{r}) G_0^{-1}(\bold{r},\bold{r}^{\prime})\psi(\bold{r}^{\prime}),
\end{equation}
where
\begin{equation}
G_{0}^{-1}(\bold{r},\bold{r}^{\prime})=-\nabla \varepsilon(\bold{r})\nabla \delta(\bold{r}-\bold{r}^{\prime})
\end{equation}
with the local permittivity
\begin{equation}
\varepsilon(\bold{r})=\varepsilon(z)=
\begin{cases}
\varepsilon_{m}, & z<0\,\\
\varepsilon_{s},&z>0,
\end{cases}
\end{equation}

Given that external potentials near the interface only affect ions within a rather thin layer, we can make the following "mean-field" approximation~\cite{budkov2022modified}:
\begin{equation}
e^{-\beta u_{\alpha}(z)}-1\approx b_{\alpha}\delta(z),
\end{equation}so that we obtain
\begin{multline}
W[G;\psi]=\frac{1}{2}\ln\frac{Det G}{Det G_{0}}-\frac{1}{2}tr\left(G[G_0^{-1}-G^{-1}]\right)+\frac{\beta}{2}(\psi G_0^{-1}\psi)+\\\sum\limits_{\alpha}z_{\alpha}\int d\bold{r}\theta(z)A_{\alpha}(z)e^{-\beta q_{\alpha}\psi(\bold{r})}+\sum\limits_{\alpha}z_{\alpha}b_{\alpha}\int d\bold{r}A_{\alpha}(0)e^{-\beta q_{\alpha}\psi(\bold{r})}\delta(z),
\end{multline}
where $b_{\alpha}$ represents the values with a dimension of length, and is determined by the integrals
\begin{equation}
b_{\alpha}=\int\limits_{0}^{\infty}dz\left(e^{-\beta u_{\alpha}(z)}-1\right),
\end{equation}
The variables $b_{\alpha}$, which generally are the functions of temperature, can be seen as measures of the adhesion of ions to the interface. Note that following the Gibbs adsorption theory we essentially consider the concept of an infinitely thin adsorption layer composed of ions~\cite{derjaguin1987derjaguin}.

Thus varying the functional $W$ with respect to $\psi$ and $G$, i.e. considering the following Euler-Lagrange equations
\begin{equation}
\frac{\delta W}{\delta \psi(\bold{r})}=0,~\frac{\delta W}{\delta G(\bold{r},\bold{r}')}=0,
\end{equation}
using the 2D Fourier-Bessel transform for the trial Green function
\begin{equation}\label{}
G(\bold{r},\bold{r}')=G(\bm{\rho}-\bm{\rho}';z,z')=\int \frac{d^2\bold{q}}{(2\pi)^2}e^{-i\bold{q}(\bm{\rho}-\bm{\rho}')}G(q|z,z'),
\end{equation}
and taking into account that $\psi(\bold{r})=\psi(z)$, we arrive at the following self-consistent field equations
\begin{equation}
\label{scf_eq2}
\partial_{z}\varepsilon(z)\partial_{z}\psi(z)=-\sum\limits_{\alpha}q_{\alpha}\bar{n}_{\alpha}(z)-\sigma_s\delta(z),
\end{equation}
\begin{equation}
\label{scf_eq1}
\left(-\partial_{z}\varepsilon(z)\partial_{z}+\varepsilon(z)(q^2+\varkappa^2(z))+\varepsilon_{s}\kappa_s\delta(z)\right)G(q|z,z')=\delta(z-z'),
\end{equation}
where we have introduced the average ionic concentrations~\cite{budkov2023variational}
\begin{equation}
\label{n_alpha}
\bar{n}_{\alpha}(z)=z_{\alpha}A_{\alpha}(z)e^{-\beta q_{\alpha}\psi(z)}\theta(z),
\end{equation}
the auxiliary functions
\begin{equation}
A_{\alpha}(z)=\exp\left[-\frac{\beta q_{\alpha}^2 G(0;z,z)}{2}\right],
\end{equation}
and the following short-hand notations
\begin{equation}
\varkappa^2(z)=\frac{1}{\varepsilon_s}\sum\limits_{\alpha}q_{\alpha}^2z_{\alpha} A_{\alpha}(z)e^{-\beta q_{\alpha}\psi(z)}\theta(z),
\end{equation}
\begin{equation}
\kappa_s=\frac{1}{\varepsilon_s}\sum\limits_{\alpha}q_{\alpha}^2z_{\alpha}b_{\alpha}A_{\alpha}(0)e^{-\beta q_{\alpha}\psi(0)}.
\end{equation}
The effective surface charge density of the interface that is appeared due to adsorption of ions is
\begin{equation}
\sigma_{s}=\sum\limits_{\alpha}q_{\alpha}z_{\alpha}b_{\alpha}A_{\alpha}(0)e^{-\beta q_{\alpha}\psi(0)}.
\end{equation}

In the mean-field approximation we have $A_{\alpha}(z)\approx 1$ and in the regime of weak electrostatic interactions, eqs. (\ref{scf_eq2}) and (\ref{scf_eq1}) transform into the following 
\begin{equation}
\label{lin_PB}
\psi^{\prime\prime}(z)-\varkappa^2(z)\psi(z) = 0,
\end{equation}
\begin{equation}
\label{eq_for_G}
\left(-\partial_{z}\varepsilon(z)\partial_{z}+\varepsilon(z)(q^2+\varkappa^2(z))+\varepsilon_{s}\kappa_s\delta(z)\right)G(q|z,z')=\delta(z-z'),
\end{equation}
where
\begin{equation}
\varkappa^2(z)\approx \kappa^2\theta(z),~\kappa^2=\frac{1}{\varepsilon_s k_{B}T}\sum\limits_{\alpha}q_{\alpha}^2 n_{\alpha},
\end{equation}
\begin{equation}
\kappa_s=\frac{1}{\varepsilon_s}\sum\limits_{\alpha}q_{\alpha}^2n_{\alpha}b_{\alpha}(1-\beta q_{\alpha}\psi_0),
\end{equation}
and we took into account that in the mean-field approximation $z_{\alpha}=n_{\alpha}$. The equation (\ref{lin_PB}) represents the standard linearized PB equation for $z>0$ and the Laplace equation for $z<0$. Note that eqs. (\ref{lin_PB}) and (\ref{eq_for_G}) are used in the main text.

\bibliography{name}

\begin{thebibliography}{53}%
\makeatletter
\providecommand \@ifxundefined [1]{%
 \@ifx{#1\undefined}
}%
\providecommand \@ifnum [1]{%
 \ifnum #1\expandafter \@firstoftwo
 \else \expandafter \@secondoftwo
 \fi
}%
\providecommand \@ifx [1]{%
 \ifx #1\expandafter \@firstoftwo
 \else \expandafter \@secondoftwo
 \fi
}%
\providecommand \natexlab [1]{#1}%
\providecommand \enquote  [1]{``#1''}%
\providecommand \bibnamefont  [1]{#1}%
\providecommand \bibfnamefont [1]{#1}%
\providecommand \citenamefont [1]{#1}%
\providecommand \href@noop [0]{\@secondoftwo}%
\providecommand \href [0]{\begingroup \@sanitize@url \@href}%
\providecommand \@href[1]{\@@startlink{#1}\@@href}%
\providecommand \@@href[1]{\endgroup#1\@@endlink}%
\providecommand \@sanitize@url [0]{\catcode `\\12\catcode `\$12\catcode `\&12\catcode `\#12\catcode `\^12\catcode `\_12\catcode `\%12\relax}%
\providecommand \@@startlink[1]{}%
\providecommand \@@endlink[0]{}%
\providecommand \url  [0]{\begingroup\@sanitize@url \@url }%
\providecommand \@url [1]{\endgroup\@href {#1}{\urlprefix }}%
\providecommand \urlprefix  [0]{URL }%
\providecommand \Eprint [0]{\href }%
\providecommand \doibase [0]{https://doi.org/}%
\providecommand \selectlanguage [0]{\@gobble}%
\providecommand \bibinfo  [0]{\@secondoftwo}%
\providecommand \bibfield  [0]{\@secondoftwo}%
\providecommand \translation [1]{[#1]}%
\providecommand \BibitemOpen [0]{}%
\providecommand \bibitemStop [0]{}%
\providecommand \bibitemNoStop [0]{.\EOS\space}%
\providecommand \EOS [0]{\spacefactor3000\relax}%
\providecommand \BibitemShut  [1]{\csname bibitem#1\endcsname}%
\let\auto@bib@innerbib\@empty
\bibitem [{\citenamefont {Wagner}(1924)}]{wagner1924oberflachenspannung}%
  \BibitemOpen
  \bibfield  {author} {\bibinfo {author} {\bibfnamefont {C.}~\bibnamefont {Wagner}},\ }\bibfield  {title} {\enquote {\bibinfo {title} {Die oberfl{\"a}chenspannung verd{\"u}nnter elektrolytl{\"o}sungen},}\ }\href@noop {} {\bibfield  {journal} {\bibinfo  {journal} {Phys. Z}\ }\textbf {\bibinfo {volume} {25}},\ \bibinfo {pages} {474--477} (\bibinfo {year} {1924})}\BibitemShut {NoStop}%
\bibitem [{\citenamefont {Onsager}\ and\ \citenamefont {Samaras}(1934)}]{onsager1934surface}%
  \BibitemOpen
  \bibfield  {author} {\bibinfo {author} {\bibfnamefont {L.}~\bibnamefont {Onsager}}\ and\ \bibinfo {author} {\bibfnamefont {N.~N.}\ \bibnamefont {Samaras}},\ }\bibfield  {title} {\enquote {\bibinfo {title} {The surface tension of debye-h{\"u}ckel electrolytes},}\ }\href@noop {} {\bibfield  {journal} {\bibinfo  {journal} {The Journal of Chemical Physics}\ }\textbf {\bibinfo {volume} {2}},\ \bibinfo {pages} {528--536} (\bibinfo {year} {1934})}\BibitemShut {NoStop}%
\bibitem [{\citenamefont {Landau}\ \emph {et~al.}(2013)\citenamefont {Landau}, \citenamefont {Bell}, \citenamefont {Kearsley}, \citenamefont {Pitaevskii}, \citenamefont {Lifshitz},\ and\ \citenamefont {Sykes}}]{landau2013electrodynamics}%
  \BibitemOpen
  \bibfield  {author} {\bibinfo {author} {\bibfnamefont {L.~D.}\ \bibnamefont {Landau}}, \bibinfo {author} {\bibfnamefont {J.~S.}\ \bibnamefont {Bell}}, \bibinfo {author} {\bibfnamefont {M.}~\bibnamefont {Kearsley}}, \bibinfo {author} {\bibfnamefont {L.}~\bibnamefont {Pitaevskii}}, \bibinfo {author} {\bibfnamefont {E.}~\bibnamefont {Lifshitz}},\ and\ \bibinfo {author} {\bibfnamefont {J.}~\bibnamefont {Sykes}},\ }\href@noop {} {\emph {\bibinfo {title} {Electrodynamics of continuous media}}},\ Vol.~\bibinfo {volume} {8}\ (\bibinfo  {publisher} {elsevier},\ \bibinfo {year} {2013})\BibitemShut {NoStop}%
\bibitem [{\citenamefont {Mucha}\ \emph {et~al.}(2005)\citenamefont {Mucha}, \citenamefont {Frigato}, \citenamefont {Levering}, \citenamefont {Allen}, \citenamefont {Tobias}, \citenamefont {Dang},\ and\ \citenamefont {Jungwirth}}]{mucha2005unified}%
  \BibitemOpen
  \bibfield  {author} {\bibinfo {author} {\bibfnamefont {M.}~\bibnamefont {Mucha}}, \bibinfo {author} {\bibfnamefont {T.}~\bibnamefont {Frigato}}, \bibinfo {author} {\bibfnamefont {L.~M.}\ \bibnamefont {Levering}}, \bibinfo {author} {\bibfnamefont {H.~C.}\ \bibnamefont {Allen}}, \bibinfo {author} {\bibfnamefont {D.~J.}\ \bibnamefont {Tobias}}, \bibinfo {author} {\bibfnamefont {L.~X.}\ \bibnamefont {Dang}},\ and\ \bibinfo {author} {\bibfnamefont {P.}~\bibnamefont {Jungwirth}},\ }\href@noop {} {\enquote {\bibinfo {title} {Unified molecular picture of the surfaces of aqueous acid, base, and salt solutions},}\ } (\bibinfo {year} {2005})\BibitemShut {NoStop}%
\bibitem [{\citenamefont {Fennell}\ \emph {et~al.}(2009)\citenamefont {Fennell}, \citenamefont {Bizjak}, \citenamefont {Vlachy},\ and\ \citenamefont {Dill}}]{fennell2009ion}%
  \BibitemOpen
  \bibfield  {author} {\bibinfo {author} {\bibfnamefont {C.~J.}\ \bibnamefont {Fennell}}, \bibinfo {author} {\bibfnamefont {A.}~\bibnamefont {Bizjak}}, \bibinfo {author} {\bibfnamefont {V.}~\bibnamefont {Vlachy}},\ and\ \bibinfo {author} {\bibfnamefont {K.~A.}\ \bibnamefont {Dill}},\ }\bibfield  {title} {\enquote {\bibinfo {title} {Ion pairing in molecular simulations of aqueous alkali halide solutions},}\ }\href@noop {} {\bibfield  {journal} {\bibinfo  {journal} {The Journal of Physical Chemistry B}\ }\textbf {\bibinfo {volume} {113}},\ \bibinfo {pages} {6782--6791} (\bibinfo {year} {2009})}\BibitemShut {NoStop}%
\bibitem [{\citenamefont {Jungwirth}\ and\ \citenamefont {Tobias}(2002)}]{jungwirth2002ions}%
  \BibitemOpen
  \bibfield  {author} {\bibinfo {author} {\bibfnamefont {P.}~\bibnamefont {Jungwirth}}\ and\ \bibinfo {author} {\bibfnamefont {D.~J.}\ \bibnamefont {Tobias}},\ }\href@noop {} {\enquote {\bibinfo {title} {Ions at the air/water interface},}\ } (\bibinfo {year} {2002})\BibitemShut {NoStop}%
\bibitem [{\citenamefont {Shchekin}\ and\ \citenamefont {Borisov}(2005)}]{shchekin2005thermodynamics}%
  \BibitemOpen
  \bibfield  {author} {\bibinfo {author} {\bibfnamefont {A.}~\bibnamefont {Shchekin}}\ and\ \bibinfo {author} {\bibfnamefont {V.}~\bibnamefont {Borisov}},\ }\bibfield  {title} {\enquote {\bibinfo {title} {Thermodynamics of nucleation on the particles of salts-strong electrolytes: the allowance for ion adsorption in the droplet surface layer},}\ }\href@noop {} {\bibfield  {journal} {\bibinfo  {journal} {Colloid Journal}\ }\textbf {\bibinfo {volume} {67}},\ \bibinfo {pages} {774--787} (\bibinfo {year} {2005})}\BibitemShut {NoStop}%
\bibitem [{\citenamefont {Ninham}\ and\ \citenamefont {Yaminsky}(1997)}]{ninham1997ion}%
  \BibitemOpen
  \bibfield  {author} {\bibinfo {author} {\bibfnamefont {B.~W.}\ \bibnamefont {Ninham}}\ and\ \bibinfo {author} {\bibfnamefont {V.}~\bibnamefont {Yaminsky}},\ }\bibfield  {title} {\enquote {\bibinfo {title} {Ion binding and ion specificity: the hofmeister effect and onsager and lifshitz theories},}\ }\href@noop {} {\bibfield  {journal} {\bibinfo  {journal} {Langmuir}\ }\textbf {\bibinfo {volume} {13}},\ \bibinfo {pages} {2097--2108} (\bibinfo {year} {1997})}\BibitemShut {NoStop}%
\bibitem [{\citenamefont {Ninham}, \citenamefont {Kurihara},\ and\ \citenamefont {Vinogradova}(1997)}]{ninham1997hydrophobicity}%
  \BibitemOpen
  \bibfield  {author} {\bibinfo {author} {\bibfnamefont {B.}~\bibnamefont {Ninham}}, \bibinfo {author} {\bibfnamefont {K.}~\bibnamefont {Kurihara}},\ and\ \bibinfo {author} {\bibfnamefont {O.}~\bibnamefont {Vinogradova}},\ }\bibfield  {title} {\enquote {\bibinfo {title} {Hydrophobicity, specific ion adsorption and reactivity},}\ }\href@noop {} {\bibfield  {journal} {\bibinfo  {journal} {Colloids and Surfaces A: Physicochemical and Engineering Aspects}\ }\textbf {\bibinfo {volume} {123}},\ \bibinfo {pages} {7--12} (\bibinfo {year} {1997})}\BibitemShut {NoStop}%
\bibitem [{\citenamefont {Levin}(2009)}]{levin2009polarizable}%
  \BibitemOpen
  \bibfield  {author} {\bibinfo {author} {\bibfnamefont {Y.}~\bibnamefont {Levin}},\ }\bibfield  {title} {\enquote {\bibinfo {title} {Polarizable ions at interfaces},}\ }\href@noop {} {\bibfield  {journal} {\bibinfo  {journal} {Physical review letters}\ }\textbf {\bibinfo {volume} {102}},\ \bibinfo {pages} {147803} (\bibinfo {year} {2009})}\BibitemShut {NoStop}%
\bibitem [{\citenamefont {Levin}, \citenamefont {Dos~Santos},\ and\ \citenamefont {Diehl}(2009)}]{levin2009ions}%
  \BibitemOpen
  \bibfield  {author} {\bibinfo {author} {\bibfnamefont {Y.}~\bibnamefont {Levin}}, \bibinfo {author} {\bibfnamefont {A.~P.}\ \bibnamefont {Dos~Santos}},\ and\ \bibinfo {author} {\bibfnamefont {A.}~\bibnamefont {Diehl}},\ }\bibfield  {title} {\enquote {\bibinfo {title} {Ions at the air-water interface: an end to a hundred-year-old mystery?}}\ }\href@noop {} {\bibfield  {journal} {\bibinfo  {journal} {Physical review letters}\ }\textbf {\bibinfo {volume} {103}},\ \bibinfo {pages} {257802} (\bibinfo {year} {2009})}\BibitemShut {NoStop}%
\bibitem [{\citenamefont {dos Santos}, \citenamefont {Diehl},\ and\ \citenamefont {Levin}(2010)}]{dos2010surface}%
  \BibitemOpen
  \bibfield  {author} {\bibinfo {author} {\bibfnamefont {A.~P.}\ \bibnamefont {dos Santos}}, \bibinfo {author} {\bibfnamefont {A.}~\bibnamefont {Diehl}},\ and\ \bibinfo {author} {\bibfnamefont {Y.}~\bibnamefont {Levin}},\ }\bibfield  {title} {\enquote {\bibinfo {title} {Surface tensions, surface potentials, and the hofmeister series of electrolyte solutions},}\ }\href@noop {} {\bibfield  {journal} {\bibinfo  {journal} {Langmuir}\ }\textbf {\bibinfo {volume} {26}},\ \bibinfo {pages} {10778--10783} (\bibinfo {year} {2010})}\BibitemShut {NoStop}%
\bibitem [{\citenamefont {dos Santos}\ and\ \citenamefont {Levin}(2012)}]{dos2012ions}%
  \BibitemOpen
  \bibfield  {author} {\bibinfo {author} {\bibfnamefont {A.~P.}\ \bibnamefont {dos Santos}}\ and\ \bibinfo {author} {\bibfnamefont {Y.}~\bibnamefont {Levin}},\ }\bibfield  {title} {\enquote {\bibinfo {title} {Ions at the water--oil interface: interfacial tension of electrolyte solutions},}\ }\href@noop {} {\bibfield  {journal} {\bibinfo  {journal} {Langmuir}\ }\textbf {\bibinfo {volume} {28}},\ \bibinfo {pages} {1304--1308} (\bibinfo {year} {2012})}\BibitemShut {NoStop}%
\bibitem [{\citenamefont {Markovich}, \citenamefont {Andelman},\ and\ \citenamefont {Podgornik}(2014)}]{markovich2014surface}%
  \BibitemOpen
  \bibfield  {author} {\bibinfo {author} {\bibfnamefont {T.}~\bibnamefont {Markovich}}, \bibinfo {author} {\bibfnamefont {D.}~\bibnamefont {Andelman}},\ and\ \bibinfo {author} {\bibfnamefont {R.}~\bibnamefont {Podgornik}},\ }\bibfield  {title} {\enquote {\bibinfo {title} {Surface tension of electrolyte solutions: A self-consistent theory},}\ }\href@noop {} {\bibfield  {journal} {\bibinfo  {journal} {Europhysics Letters}\ }\textbf {\bibinfo {volume} {106}},\ \bibinfo {pages} {16002} (\bibinfo {year} {2014})}\BibitemShut {NoStop}%
\bibitem [{\citenamefont {Markovich}, \citenamefont {Andelman},\ and\ \citenamefont {Podgornik}(2015)}]{markovich2015surface}%
  \BibitemOpen
  \bibfield  {author} {\bibinfo {author} {\bibfnamefont {T.}~\bibnamefont {Markovich}}, \bibinfo {author} {\bibfnamefont {D.}~\bibnamefont {Andelman}},\ and\ \bibinfo {author} {\bibfnamefont {R.}~\bibnamefont {Podgornik}},\ }\bibfield  {title} {\enquote {\bibinfo {title} {Surface tension of electrolyte interfaces: Ionic specificity within a field-theory approach},}\ }\href@noop {} {\bibfield  {journal} {\bibinfo  {journal} {The Journal of Chemical Physics}\ }\textbf {\bibinfo {volume} {142}} (\bibinfo {year} {2015})}\BibitemShut {NoStop}%
\bibitem [{\citenamefont {Lau}\ and\ \citenamefont {Sokoloff}(2020)}]{Lau2020}%
  \BibitemOpen
  \bibfield  {author} {\bibinfo {author} {\bibfnamefont {A.}~\bibnamefont {Lau}}\ and\ \bibinfo {author} {\bibfnamefont {J.}~\bibnamefont {Sokoloff}},\ }\bibfield  {title} {\enquote {\bibinfo {title} {Enhancement of the ion concentration in a salt solution near a wall due to electrical image potentials and enhancement of surface tension due to the presence of salt},}\ }\href@noop {} {\bibfield  {journal} {\bibinfo  {journal} {Physical Review E}\ }\textbf {\bibinfo {volume} {102}},\ \bibinfo {pages} {052606} (\bibinfo {year} {2020})}\BibitemShut {NoStop}%
\bibitem [{\citenamefont {Hatlo}\ and\ \citenamefont {Lue}(2008)}]{hatlo2008role}%
  \BibitemOpen
  \bibfield  {author} {\bibinfo {author} {\bibfnamefont {M.~M.}\ \bibnamefont {Hatlo}}\ and\ \bibinfo {author} {\bibfnamefont {L.}~\bibnamefont {Lue}},\ }\bibfield  {title} {\enquote {\bibinfo {title} {The role of image charges in the interactions between colloidal particles},}\ }\href@noop {} {\bibfield  {journal} {\bibinfo  {journal} {Soft Matter}\ }\textbf {\bibinfo {volume} {4}},\ \bibinfo {pages} {1582--1596} (\bibinfo {year} {2008})}\BibitemShut {NoStop}%
\bibitem [{\citenamefont {Wang}\ and\ \citenamefont {Wang}(2016)}]{wang2016inhomogeneous}%
  \BibitemOpen
  \bibfield  {author} {\bibinfo {author} {\bibfnamefont {R.}~\bibnamefont {Wang}}\ and\ \bibinfo {author} {\bibfnamefont {Z.-G.}\ \bibnamefont {Wang}},\ }\bibfield  {title} {\enquote {\bibinfo {title} {Inhomogeneous screening near the dielectric interface},}\ }\href@noop {} {\bibfield  {journal} {\bibinfo  {journal} {The Journal of chemical physics}\ }\textbf {\bibinfo {volume} {144}} (\bibinfo {year} {2016})}\BibitemShut {NoStop}%
\bibitem [{\citenamefont {Dean}\ and\ \citenamefont {Horgan}(2003)}]{dean2003weak}%
  \BibitemOpen
  \bibfield  {author} {\bibinfo {author} {\bibfnamefont {D.~S.}\ \bibnamefont {Dean}}\ and\ \bibinfo {author} {\bibfnamefont {R.~R.}\ \bibnamefont {Horgan}},\ }\bibfield  {title} {\enquote {\bibinfo {title} {Weak nonlinear surface-charging effects in electrolytic films},}\ }\href@noop {} {\bibfield  {journal} {\bibinfo  {journal} {Physical Review E}\ }\textbf {\bibinfo {volume} {68}},\ \bibinfo {pages} {051104} (\bibinfo {year} {2003})}\BibitemShut {NoStop}%
\bibitem [{\citenamefont {Dean}\ and\ \citenamefont {Horgan}(2004)}]{dean2004field}%
  \BibitemOpen
  \bibfield  {author} {\bibinfo {author} {\bibfnamefont {D.~S.}\ \bibnamefont {Dean}}\ and\ \bibinfo {author} {\bibfnamefont {R.~R.}\ \bibnamefont {Horgan}},\ }\bibfield  {title} {\enquote {\bibinfo {title} {Field theoretic calculation of the surface tension for a model electrolyte system},}\ }\href@noop {} {\bibfield  {journal} {\bibinfo  {journal} {Physical Review E}\ }\textbf {\bibinfo {volume} {69}},\ \bibinfo {pages} {061603} (\bibinfo {year} {2004})}\BibitemShut {NoStop}%
\bibitem [{\citenamefont {Rowlinson}\ and\ \citenamefont {Widom}(2013)}]{rowlinson2013molecular}%
  \BibitemOpen
  \bibfield  {author} {\bibinfo {author} {\bibfnamefont {J.~S.}\ \bibnamefont {Rowlinson}}\ and\ \bibinfo {author} {\bibfnamefont {B.}~\bibnamefont {Widom}},\ }\href@noop {} {\emph {\bibinfo {title} {Molecular theory of capillarity}}}\ (\bibinfo  {publisher} {Courier Corporation},\ \bibinfo {year} {2013})\BibitemShut {NoStop}%
\bibitem [{\citenamefont {Derjaguin}, \citenamefont {Churaev},\ and\ \citenamefont {Muller}(1987)}]{derjaguin1987derjaguin}%
  \BibitemOpen
  \bibfield  {author} {\bibinfo {author} {\bibfnamefont {B.}~\bibnamefont {Derjaguin}}, \bibinfo {author} {\bibfnamefont {N.}~\bibnamefont {Churaev}},\ and\ \bibinfo {author} {\bibfnamefont {V.}~\bibnamefont {Muller}},\ }\bibfield  {title} {\enquote {\bibinfo {title} {The derjaguin-landau-verwey-overbeek (dlvo) theory of stability of lyophobic colloids},}\ }in\ \href@noop {} {\emph {\bibinfo {booktitle} {Surface Forces}}}\ (\bibinfo  {publisher} {Springer},\ \bibinfo {year} {1987})\ pp.\ \bibinfo {pages} {293--310}\BibitemShut {NoStop}%
\bibitem [{\citenamefont {Irving}\ and\ \citenamefont {Kirkwood}(1950)}]{irving1950statistical}%
  \BibitemOpen
  \bibfield  {author} {\bibinfo {author} {\bibfnamefont {J.}~\bibnamefont {Irving}}\ and\ \bibinfo {author} {\bibfnamefont {J.~G.}\ \bibnamefont {Kirkwood}},\ }\bibfield  {title} {\enquote {\bibinfo {title} {The statistical mechanical theory of transport processes. iv. the equations of hydrodynamics},}\ }\href@noop {} {\bibfield  {journal} {\bibinfo  {journal} {The Journal of chemical physics}\ }\textbf {\bibinfo {volume} {18}},\ \bibinfo {pages} {817--829} (\bibinfo {year} {1950})}\BibitemShut {NoStop}%
\bibitem [{\citenamefont {Shi}\ \emph {et~al.}(2023)\citenamefont {Shi}, \citenamefont {Smith}, \citenamefont {Santiso},\ and\ \citenamefont {Gubbins}}]{shi2023perspective}%
  \BibitemOpen
  \bibfield  {author} {\bibinfo {author} {\bibfnamefont {K.}~\bibnamefont {Shi}}, \bibinfo {author} {\bibfnamefont {E.~R.}\ \bibnamefont {Smith}}, \bibinfo {author} {\bibfnamefont {E.~E.}\ \bibnamefont {Santiso}},\ and\ \bibinfo {author} {\bibfnamefont {K.~E.}\ \bibnamefont {Gubbins}},\ }\bibfield  {title} {\enquote {\bibinfo {title} {A perspective on the microscopic pressure (stress) tensor: History, current understanding, and future challenges},}\ }\href@noop {} {\bibfield  {journal} {\bibinfo  {journal} {The Journal of Chemical Physics}\ }\textbf {\bibinfo {volume} {158}},\ \bibinfo {pages} {040901} (\bibinfo {year} {2023})}\BibitemShut {NoStop}%
\bibitem [{\citenamefont {Rusanov}\ and\ \citenamefont {Shchekin}(2001)}]{rusanov2001condition}%
  \BibitemOpen
  \bibfield  {author} {\bibinfo {author} {\bibfnamefont {A.~I.}\ \bibnamefont {Rusanov}}\ and\ \bibinfo {author} {\bibfnamefont {A.~K.}\ \bibnamefont {Shchekin}},\ }\bibfield  {title} {\enquote {\bibinfo {title} {The condition of mechanical equilibrium for a non-spherical interface between phases with a non-diagonal stress tensor},}\ }\href@noop {} {\bibfield  {journal} {\bibinfo  {journal} {Colloids and Surfaces A: Physicochemical and Engineering Aspects}\ }\textbf {\bibinfo {volume} {192}},\ \bibinfo {pages} {357--362} (\bibinfo {year} {2001})}\BibitemShut {NoStop}%
\bibitem [{\citenamefont {Rusanov}, \citenamefont {Shchekin},\ and\ \citenamefont {Varshavskii}(2001)}]{rusanov2001three}%
  \BibitemOpen
  \bibfield  {author} {\bibinfo {author} {\bibfnamefont {A.}~\bibnamefont {Rusanov}}, \bibinfo {author} {\bibfnamefont {A.}~\bibnamefont {Shchekin}},\ and\ \bibinfo {author} {\bibfnamefont {V.}~\bibnamefont {Varshavskii}},\ }\bibfield  {title} {\enquote {\bibinfo {title} {Three-dimensional aspect of the surface tension: An approach based on the total pressure tensor},}\ }\href@noop {} {\bibfield  {journal} {\bibinfo  {journal} {Colloid Journal}\ }\textbf {\bibinfo {volume} {63}},\ \bibinfo {pages} {365--375} (\bibinfo {year} {2001})}\BibitemShut {NoStop}%
\bibitem [{\citenamefont {Budkov}\ and\ \citenamefont {Kolesnikov}(2022{\natexlab{a}})}]{budkov2022modified}%
  \BibitemOpen
  \bibfield  {author} {\bibinfo {author} {\bibfnamefont {Y.~A.}\ \bibnamefont {Budkov}}\ and\ \bibinfo {author} {\bibfnamefont {A.~L.}\ \bibnamefont {Kolesnikov}},\ }\bibfield  {title} {\enquote {\bibinfo {title} {Modified poisson--boltzmann equations and macroscopic forces in inhomogeneous ionic fluids},}\ }\href@noop {} {\bibfield  {journal} {\bibinfo  {journal} {Journal of Statistical Mechanics: Theory and Experiment}\ }\textbf {\bibinfo {volume} {2022}},\ \bibinfo {pages} {053205} (\bibinfo {year} {2022}{\natexlab{a}})}\BibitemShut {NoStop}%
\bibitem [{\citenamefont {Brandyshev}\ and\ \citenamefont {Budkov}(2023{\natexlab{a}})}]{brandyshev2023noether}%
  \BibitemOpen
  \bibfield  {author} {\bibinfo {author} {\bibfnamefont {P.~E.}\ \bibnamefont {Brandyshev}}\ and\ \bibinfo {author} {\bibfnamefont {Y.~A.}\ \bibnamefont {Budkov}},\ }\bibfield  {title} {\enquote {\bibinfo {title} {Noether's second theorem and covariant field theory of mechanical stresses in inhomogeneous ionic liquids},}\ }\href@noop {} {\bibfield  {journal} {\bibinfo  {journal} {The Journal of chemical physics}\ }\textbf {\bibinfo {volume} {158}} (\bibinfo {year} {2023}{\natexlab{a}})}\BibitemShut {NoStop}%
\bibitem [{\citenamefont {Budkov}\ and\ \citenamefont {Kalikin}(2023{\natexlab{a}})}]{budkov2023macroscopic}%
  \BibitemOpen
  \bibfield  {author} {\bibinfo {author} {\bibfnamefont {Y.~A.}\ \bibnamefont {Budkov}}\ and\ \bibinfo {author} {\bibfnamefont {N.~N.}\ \bibnamefont {Kalikin}},\ }\bibfield  {title} {\enquote {\bibinfo {title} {Macroscopic forces in inhomogeneous polyelectrolyte solutions},}\ }\href@noop {} {\bibfield  {journal} {\bibinfo  {journal} {Physical Review E}\ }\textbf {\bibinfo {volume} {107}},\ \bibinfo {pages} {024503} (\bibinfo {year} {2023}{\natexlab{a}})}\BibitemShut {NoStop}%
\bibitem [{\citenamefont {Vasileva}, \citenamefont {Mazur},\ and\ \citenamefont {Budkov}(2023)}]{vasileva2023theory}%
  \BibitemOpen
  \bibfield  {author} {\bibinfo {author} {\bibfnamefont {V.~A.}\ \bibnamefont {Vasileva}}, \bibinfo {author} {\bibfnamefont {D.~A.}\ \bibnamefont {Mazur}},\ and\ \bibinfo {author} {\bibfnamefont {Y.~A.}\ \bibnamefont {Budkov}},\ }\bibfield  {title} {\enquote {\bibinfo {title} {Theory of electrolyte solutions in a slit charged pore: effects of structural interactions and specific adsorption of ions},}\ }\href@noop {} {\bibfield  {journal} {\bibinfo  {journal} {The Journal of Chemical Physics}\ }\textbf {\bibinfo {volume} {159}} (\bibinfo {year} {2023})}\BibitemShut {NoStop}%
\bibitem [{\citenamefont {Budkov}\ and\ \citenamefont {Kalikin}(2023{\natexlab{b}})}]{budkov2023dielectric}%
  \BibitemOpen
  \bibfield  {author} {\bibinfo {author} {\bibfnamefont {Y.~A.}\ \bibnamefont {Budkov}}\ and\ \bibinfo {author} {\bibfnamefont {N.~N.}\ \bibnamefont {Kalikin}},\ }\bibfield  {title} {\enquote {\bibinfo {title} {Dielectric mismatch effects on polyelectrolyte solutions in electrified nanopores: Insights from mean-field theory},}\ }\href@noop {} {\bibfield  {journal} {\bibinfo  {journal} {Polym. Sci. Ser. C}\ }\textbf {\bibinfo {volume} {65}},\ \bibinfo {pages} {46} (\bibinfo {year} {2023}{\natexlab{b}})}\BibitemShut {NoStop}%
\bibitem [{\citenamefont {Budkov}\ and\ \citenamefont {Brandyshev}(2023)}]{budkov2023variational}%
  \BibitemOpen
  \bibfield  {author} {\bibinfo {author} {\bibfnamefont {Y.~A.}\ \bibnamefont {Budkov}}\ and\ \bibinfo {author} {\bibfnamefont {P.~E.}\ \bibnamefont {Brandyshev}},\ }\bibfield  {title} {\enquote {\bibinfo {title} {Variational field theory of macroscopic forces in coulomb fluids},}\ }\href@noop {} {\bibfield  {journal} {\bibinfo  {journal} {The Journal of Chemical Physics}\ }\textbf {\bibinfo {volume} {159}} (\bibinfo {year} {2023})}\BibitemShut {NoStop}%
\bibitem [{\citenamefont {Brandyshev}\ and\ \citenamefont {Budkov}(2023{\natexlab{b}})}]{brandyshev2023statistical}%
  \BibitemOpen
  \bibfield  {author} {\bibinfo {author} {\bibfnamefont {P.~E.}\ \bibnamefont {Brandyshev}}\ and\ \bibinfo {author} {\bibfnamefont {Y.~A.}\ \bibnamefont {Budkov}},\ }\bibfield  {title} {\enquote {\bibinfo {title} {Statistical field theory of mechanical stresses in coulomb fluids: general covariant approach vs noether’s theorem},}\ }\href@noop {} {\bibfield  {journal} {\bibinfo  {journal} {Journal of Statistical Mechanics: Theory and Experiment}\ }\textbf {\bibinfo {volume} {2023}},\ \bibinfo {pages} {123206} (\bibinfo {year} {2023}{\natexlab{b}})}\BibitemShut {NoStop}%
\bibitem [{\citenamefont {Hermann}\ and\ \citenamefont {Schmidt}(2022)}]{hermann2022noether}%
  \BibitemOpen
  \bibfield  {author} {\bibinfo {author} {\bibfnamefont {S.}~\bibnamefont {Hermann}}\ and\ \bibinfo {author} {\bibfnamefont {M.}~\bibnamefont {Schmidt}},\ }\bibfield  {title} {\enquote {\bibinfo {title} {Why noether's theorem applies to statistical mechanics},}\ }\href@noop {} {\bibfield  {journal} {\bibinfo  {journal} {Journal of Physics: Condensed Matter}\ }\textbf {\bibinfo {volume} {34}},\ \bibinfo {pages} {213001} (\bibinfo {year} {2022})}\BibitemShut {NoStop}%
\bibitem [{\citenamefont {Wang}(2010)}]{wang2010fluctuation}%
  \BibitemOpen
  \bibfield  {author} {\bibinfo {author} {\bibfnamefont {Z.-G.}\ \bibnamefont {Wang}},\ }\bibfield  {title} {\enquote {\bibinfo {title} {Fluctuation in electrolyte solutions: The self energy},}\ }\href@noop {} {\bibfield  {journal} {\bibinfo  {journal} {Physical Review E}\ }\textbf {\bibinfo {volume} {81}},\ \bibinfo {pages} {021501} (\bibinfo {year} {2010})}\BibitemShut {NoStop}%
\bibitem [{\citenamefont {Dzyaloshinskii}, \citenamefont {Lifshitz},\ and\ \citenamefont {Pitaevskii}(1961)}]{dzyaloshinskii1961general}%
  \BibitemOpen
  \bibfield  {author} {\bibinfo {author} {\bibfnamefont {I.~E.}\ \bibnamefont {Dzyaloshinskii}}, \bibinfo {author} {\bibfnamefont {E.~M.}\ \bibnamefont {Lifshitz}},\ and\ \bibinfo {author} {\bibfnamefont {L.~P.}\ \bibnamefont {Pitaevskii}},\ }\bibfield  {title} {\enquote {\bibinfo {title} {The general theory of van der waals forces},}\ }\href@noop {} {\bibfield  {journal} {\bibinfo  {journal} {Advances in Physics}\ }\textbf {\bibinfo {volume} {10}},\ \bibinfo {pages} {165--209} (\bibinfo {year} {1961})}\BibitemShut {NoStop}%
\bibitem [{\citenamefont {Markovich}, \citenamefont {Andelman},\ and\ \citenamefont {Podgornik}(2017)}]{markovich2017surface}%
  \BibitemOpen
  \bibfield  {author} {\bibinfo {author} {\bibfnamefont {T.}~\bibnamefont {Markovich}}, \bibinfo {author} {\bibfnamefont {D.}~\bibnamefont {Andelman}},\ and\ \bibinfo {author} {\bibfnamefont {R.}~\bibnamefont {Podgornik}},\ }\bibfield  {title} {\enquote {\bibinfo {title} {Surface tension of acid solutions: Fluctuations beyond the nonlinear poisson--boltzmann theory},}\ }\href@noop {} {\bibfield  {journal} {\bibinfo  {journal} {Langmuir}\ }\textbf {\bibinfo {volume} {33}},\ \bibinfo {pages} {34--44} (\bibinfo {year} {2017})}\BibitemShut {NoStop}%
\bibitem [{\citenamefont {Nightingale~Jr}(1959)}]{nightingale1959phenomenological}%
  \BibitemOpen
  \bibfield  {author} {\bibinfo {author} {\bibfnamefont {E.}~\bibnamefont {Nightingale~Jr}},\ }\bibfield  {title} {\enquote {\bibinfo {title} {Phenomenological theory of ion solvation. effective radii of hydrated ions},}\ }\href@noop {} {\bibfield  {journal} {\bibinfo  {journal} {The Journal of Physical Chemistry}\ }\textbf {\bibinfo {volume} {63}},\ \bibinfo {pages} {1381--1387} (\bibinfo {year} {1959})}\BibitemShut {NoStop}%
\bibitem [{\citenamefont {Matubayasi}(2013)}]{matubayasi2013surface}%
  \BibitemOpen
  \bibfield  {author} {\bibinfo {author} {\bibfnamefont {N.}~\bibnamefont {Matubayasi}},\ }\href@noop {} {\emph {\bibinfo {title} {Surface tension and related thermodynamic quantities of aqueous electrolyte solutions}}}\ (\bibinfo  {publisher} {CRC Press},\ \bibinfo {year} {2013})\BibitemShut {NoStop}%
\bibitem [{\citenamefont {Aveyard}\ and\ \citenamefont {Saleem}(1976)}]{aveyard1976interfacial}%
  \BibitemOpen
  \bibfield  {author} {\bibinfo {author} {\bibfnamefont {R.}~\bibnamefont {Aveyard}}\ and\ \bibinfo {author} {\bibfnamefont {S.~M.}\ \bibnamefont {Saleem}},\ }\bibfield  {title} {\enquote {\bibinfo {title} {Interfacial tensions at alkane-aqueous electrolyte interfaces},}\ }\href@noop {} {\bibfield  {journal} {\bibinfo  {journal} {Journal of the Chemical Society, Faraday Transactions 1: Physical Chemistry in Condensed Phases}\ }\textbf {\bibinfo {volume} {72}},\ \bibinfo {pages} {1609--1617} (\bibinfo {year} {1976})}\BibitemShut {NoStop}%
\bibitem [{\citenamefont {Wang}\ and\ \citenamefont {Wang}(2015)}]{wang2015theoretical}%
  \BibitemOpen
  \bibfield  {author} {\bibinfo {author} {\bibfnamefont {R.}~\bibnamefont {Wang}}\ and\ \bibinfo {author} {\bibfnamefont {Z.-G.}\ \bibnamefont {Wang}},\ }\bibfield  {title} {\enquote {\bibinfo {title} {On the theoretical description of weakly charged surfaces},}\ }\href@noop {} {\bibfield  {journal} {\bibinfo  {journal} {The Journal of Chemical Physics}\ }\textbf {\bibinfo {volume} {142}} (\bibinfo {year} {2015})}\BibitemShut {NoStop}%
\bibitem [{\citenamefont {Urbanija}\ \emph {et~al.}(2008)\citenamefont {Urbanija}, \citenamefont {Bohinc}, \citenamefont {Bellen}, \citenamefont {Maset}, \citenamefont {Igli{\v{c}}}, \citenamefont {Kralj-Igli{\v{c}}},\ and\ \citenamefont {Sunil~Kumar}}]{urbanija2008attraction}%
  \BibitemOpen
  \bibfield  {author} {\bibinfo {author} {\bibfnamefont {J.}~\bibnamefont {Urbanija}}, \bibinfo {author} {\bibfnamefont {K.}~\bibnamefont {Bohinc}}, \bibinfo {author} {\bibfnamefont {A.}~\bibnamefont {Bellen}}, \bibinfo {author} {\bibfnamefont {S.}~\bibnamefont {Maset}}, \bibinfo {author} {\bibfnamefont {A.}~\bibnamefont {Igli{\v{c}}}}, \bibinfo {author} {\bibfnamefont {V.}~\bibnamefont {Kralj-Igli{\v{c}}}},\ and\ \bibinfo {author} {\bibfnamefont {P.}~\bibnamefont {Sunil~Kumar}},\ }\bibfield  {title} {\enquote {\bibinfo {title} {Attraction between negatively charged surfaces mediated by spherical counterions with quadrupolar charge distribution},}\ }\href@noop {} {\bibfield  {journal} {\bibinfo  {journal} {The Journal of chemical physics}\ }\textbf {\bibinfo {volume} {129}} (\bibinfo {year} {2008})}\BibitemShut {NoStop}%
\bibitem [{\citenamefont {Kondrat}\ \emph {et~al.}(2023)\citenamefont {Kondrat}, \citenamefont {Feng}, \citenamefont {Bresme}, \citenamefont {Urbakh},\ and\ \citenamefont {Kornyshev}}]{kondrat2023theory}%
  \BibitemOpen
  \bibfield  {author} {\bibinfo {author} {\bibfnamefont {S.}~\bibnamefont {Kondrat}}, \bibinfo {author} {\bibfnamefont {G.}~\bibnamefont {Feng}}, \bibinfo {author} {\bibfnamefont {F.}~\bibnamefont {Bresme}}, \bibinfo {author} {\bibfnamefont {M.}~\bibnamefont {Urbakh}},\ and\ \bibinfo {author} {\bibfnamefont {A.~A.}\ \bibnamefont {Kornyshev}},\ }\bibfield  {title} {\enquote {\bibinfo {title} {Theory and simulations of ionic liquids in nanoconfinement},}\ }\href@noop {} {\bibfield  {journal} {\bibinfo  {journal} {Chemical Reviews}\ }\textbf {\bibinfo {volume} {123}},\ \bibinfo {pages} {6668--6715} (\bibinfo {year} {2023})}\BibitemShut {NoStop}%
\bibitem [{\citenamefont {Budkov}\ and\ \citenamefont {Kolesnikov}(2022{\natexlab{b}})}]{budkov2022electric}%
  \BibitemOpen
  \bibfield  {author} {\bibinfo {author} {\bibfnamefont {Y.~A.}\ \bibnamefont {Budkov}}\ and\ \bibinfo {author} {\bibfnamefont {A.~L.}\ \bibnamefont {Kolesnikov}},\ }\bibfield  {title} {\enquote {\bibinfo {title} {Electric double layer theory for room temperature ionic liquids on charged electrodes: Milestones and prospects},}\ }\href@noop {} {\bibfield  {journal} {\bibinfo  {journal} {Current Opinion in Electrochemistry}\ }\textbf {\bibinfo {volume} {33}},\ \bibinfo {pages} {100931} (\bibinfo {year} {2022}{\natexlab{b}})}\BibitemShut {NoStop}%
\bibitem [{\citenamefont {Gongadze}\ and\ \citenamefont {Igli{\v{c}}}(2012)}]{gongadze2012decrease}%
  \BibitemOpen
  \bibfield  {author} {\bibinfo {author} {\bibfnamefont {E.}~\bibnamefont {Gongadze}}\ and\ \bibinfo {author} {\bibfnamefont {A.}~\bibnamefont {Igli{\v{c}}}},\ }\bibfield  {title} {\enquote {\bibinfo {title} {Decrease of permittivity of an electrolyte solution near a charged surface due to saturation and excluded volume effects},}\ }\href@noop {} {\bibfield  {journal} {\bibinfo  {journal} {Bioelectrochemistry}\ }\textbf {\bibinfo {volume} {87}},\ \bibinfo {pages} {199--203} (\bibinfo {year} {2012})}\BibitemShut {NoStop}%
\bibitem [{\citenamefont {Gongadze}\ \emph {et~al.}(2014)\citenamefont {Gongadze}, \citenamefont {Velikonja}, \citenamefont {Perutkova}, \citenamefont {Kramar}, \citenamefont {Ma{\v{c}}ek-Lebar}, \citenamefont {Kralj-Igli{\v{c}}},\ and\ \citenamefont {Igli{\v{c}}}}]{gongadze2014ions}%
  \BibitemOpen
  \bibfield  {author} {\bibinfo {author} {\bibfnamefont {E.}~\bibnamefont {Gongadze}}, \bibinfo {author} {\bibfnamefont {A.}~\bibnamefont {Velikonja}}, \bibinfo {author} {\bibfnamefont {{\v{S}}.}~\bibnamefont {Perutkova}}, \bibinfo {author} {\bibfnamefont {P.}~\bibnamefont {Kramar}}, \bibinfo {author} {\bibfnamefont {A.}~\bibnamefont {Ma{\v{c}}ek-Lebar}}, \bibinfo {author} {\bibfnamefont {V.}~\bibnamefont {Kralj-Igli{\v{c}}}},\ and\ \bibinfo {author} {\bibfnamefont {A.}~\bibnamefont {Igli{\v{c}}}},\ }\bibfield  {title} {\enquote {\bibinfo {title} {Ions and water molecules in an electrolyte solution in contact with charged and dipolar surfaces},}\ }\href@noop {} {\bibfield  {journal} {\bibinfo  {journal} {Electrochimica Acta}\ }\textbf {\bibinfo {volume} {126}},\ \bibinfo {pages} {42--60} (\bibinfo {year} {2014})}\BibitemShut {NoStop}%
\bibitem [{\citenamefont {Marcovitz}, \citenamefont {Naftaly},\ and\ \citenamefont {Levy}(2015)}]{marcovitz2015water}%
  \BibitemOpen
  \bibfield  {author} {\bibinfo {author} {\bibfnamefont {A.}~\bibnamefont {Marcovitz}}, \bibinfo {author} {\bibfnamefont {A.}~\bibnamefont {Naftaly}},\ and\ \bibinfo {author} {\bibfnamefont {Y.}~\bibnamefont {Levy}},\ }\bibfield  {title} {\enquote {\bibinfo {title} {Water organization between oppositely charged surfaces: Implications for protein sliding along dna},}\ }\href@noop {} {\bibfield  {journal} {\bibinfo  {journal} {The Journal of Chemical Physics}\ }\textbf {\bibinfo {volume} {142}} (\bibinfo {year} {2015})}\BibitemShut {NoStop}%
\bibitem [{\citenamefont {Budkov}\ \emph {et~al.}(2018)\citenamefont {Budkov}, \citenamefont {Kolesnikov}, \citenamefont {Goodwin}, \citenamefont {Kiselev},\ and\ \citenamefont {Kornyshev}}]{budkov2018theory}%
  \BibitemOpen
  \bibfield  {author} {\bibinfo {author} {\bibfnamefont {Y.~A.}\ \bibnamefont {Budkov}}, \bibinfo {author} {\bibfnamefont {A.~L.}\ \bibnamefont {Kolesnikov}}, \bibinfo {author} {\bibfnamefont {Z.~A.}\ \bibnamefont {Goodwin}}, \bibinfo {author} {\bibfnamefont {M.~G.}\ \bibnamefont {Kiselev}},\ and\ \bibinfo {author} {\bibfnamefont {A.~A.}\ \bibnamefont {Kornyshev}},\ }\bibfield  {title} {\enquote {\bibinfo {title} {Theory of electrosorption of water from ionic liquids},}\ }\href@noop {} {\bibfield  {journal} {\bibinfo  {journal} {Electrochimica Acta}\ }\textbf {\bibinfo {volume} {284}},\ \bibinfo {pages} {346--354} (\bibinfo {year} {2018})}\BibitemShut {NoStop}%
\bibitem [{\citenamefont {Dubtsov}\ \emph {et~al.}(2018)\citenamefont {Dubtsov}, \citenamefont {Pasechnik}, \citenamefont {Shmeliova}, \citenamefont {Saidgaziev}, \citenamefont {Gongadze}, \citenamefont {Igli{\v{c}}},\ and\ \citenamefont {Kralj}}]{dubtsov2018liquid}%
  \BibitemOpen
  \bibfield  {author} {\bibinfo {author} {\bibfnamefont {A.~V.}\ \bibnamefont {Dubtsov}}, \bibinfo {author} {\bibfnamefont {S.~V.}\ \bibnamefont {Pasechnik}}, \bibinfo {author} {\bibfnamefont {D.~V.}\ \bibnamefont {Shmeliova}}, \bibinfo {author} {\bibfnamefont {A.~S.}\ \bibnamefont {Saidgaziev}}, \bibinfo {author} {\bibfnamefont {E.}~\bibnamefont {Gongadze}}, \bibinfo {author} {\bibfnamefont {A.}~\bibnamefont {Igli{\v{c}}}},\ and\ \bibinfo {author} {\bibfnamefont {S.}~\bibnamefont {Kralj}},\ }\bibfield  {title} {\enquote {\bibinfo {title} {Liquid crystalline droplets in aqueous environments: Electrostatic effects},}\ }\href@noop {} {\bibfield  {journal} {\bibinfo  {journal} {Soft Matter}\ }\textbf {\bibinfo {volume} {14}},\ \bibinfo {pages} {9619--9630} (\bibinfo {year} {2018})}\BibitemShut {NoStop}%
\bibitem [{\citenamefont {Abrashkin}, \citenamefont {Andelman},\ and\ \citenamefont {Orland}(2007)}]{abrashkin2007dipolar}%
  \BibitemOpen
  \bibfield  {author} {\bibinfo {author} {\bibfnamefont {A.}~\bibnamefont {Abrashkin}}, \bibinfo {author} {\bibfnamefont {D.}~\bibnamefont {Andelman}},\ and\ \bibinfo {author} {\bibfnamefont {H.}~\bibnamefont {Orland}},\ }\bibfield  {title} {\enquote {\bibinfo {title} {Dipolar poisson-boltzmann equation: ions and dipoles close to charge interfaces},}\ }\href@noop {} {\bibfield  {journal} {\bibinfo  {journal} {Physical review letters}\ }\textbf {\bibinfo {volume} {99}},\ \bibinfo {pages} {077801} (\bibinfo {year} {2007})}\BibitemShut {NoStop}%
\bibitem [{\citenamefont {Cruz}\ \emph {et~al.}(2019)\citenamefont {Cruz}, \citenamefont {Kondrat}, \citenamefont {Lomba},\ and\ \citenamefont {Ciach}}]{cruz2019effect}%
  \BibitemOpen
  \bibfield  {author} {\bibinfo {author} {\bibfnamefont {C.}~\bibnamefont {Cruz}}, \bibinfo {author} {\bibfnamefont {S.}~\bibnamefont {Kondrat}}, \bibinfo {author} {\bibfnamefont {E.}~\bibnamefont {Lomba}},\ and\ \bibinfo {author} {\bibfnamefont {A.}~\bibnamefont {Ciach}},\ }\bibfield  {title} {\enquote {\bibinfo {title} {Effect of proximity to ionic liquid-solvent demixing on electrical double layers},}\ }\href@noop {} {\bibfield  {journal} {\bibinfo  {journal} {Journal of Molecular Liquids}\ }\textbf {\bibinfo {volume} {294}},\ \bibinfo {pages} {111368} (\bibinfo {year} {2019})}\BibitemShut {NoStop}%
\bibitem [{\citenamefont {Cruz}\ and\ \citenamefont {Ciach}(2021)}]{cruz2021phase}%
  \BibitemOpen
  \bibfield  {author} {\bibinfo {author} {\bibfnamefont {C.}~\bibnamefont {Cruz}}\ and\ \bibinfo {author} {\bibfnamefont {A.}~\bibnamefont {Ciach}},\ }\bibfield  {title} {\enquote {\bibinfo {title} {Phase transitions and electrochemical properties of ionic liquids and ionic liquid-solvent mixtures},}\ }\href@noop {} {\bibfield  {journal} {\bibinfo  {journal} {Molecules}\ }\textbf {\bibinfo {volume} {26}},\ \bibinfo {pages} {3668} (\bibinfo {year} {2021})}\BibitemShut {NoStop}%
\bibitem [{\citenamefont {Blossey}, \citenamefont {Maggs},\ and\ \citenamefont {Podgornik}(2017)}]{blossey2017structural}%
  \BibitemOpen
  \bibfield  {author} {\bibinfo {author} {\bibfnamefont {R.}~\bibnamefont {Blossey}}, \bibinfo {author} {\bibfnamefont {A.}~\bibnamefont {Maggs}},\ and\ \bibinfo {author} {\bibfnamefont {R.}~\bibnamefont {Podgornik}},\ }\bibfield  {title} {\enquote {\bibinfo {title} {Structural interactions in ionic liquids linked to higher-order poisson-boltzmann equations},}\ }\href@noop {} {\bibfield  {journal} {\bibinfo  {journal} {Physical Review E}\ }\textbf {\bibinfo {volume} {95}},\ \bibinfo {pages} {060602} (\bibinfo {year} {2017})}\BibitemShut {NoStop}%
\end{thebibliography}%

\end{document}